\documentclass[10pt]{iopart}
\usepackage{graphicx}
\usepackage{appendix}
\usepackage{cite}
\usepackage{ragged2e}
\usepackage{multirow}

\usepackage{dcolumn}% Align table columns on decimal point
\usepackage{bm}% bold math
\usepackage{subcaption}
\usepackage{mathtools}
\usepackage{enumerate}
\usepackage{enumitem}

\usepackage[utf8]{inputenc}
\usepackage[T1]{fontenc}

\begin{document}

\title{Perturbative model for the saturation of energetic-particle-driven modes limited by self-generated zonal modes}

\author{T Barberis$^{1}$,V N Duarte$^{1}$, E J Hartigan-O'Connor$^{1}$ and N N Gorelenkov$^{1}$}

\address{$^{1}$ Princeton Plasma Physics Laboratory, Princeton University, NJ, 08543, United States of
America}

\ead{tbarberi@pppl.gov} 
 
\begin{abstract}
We present a simplified energy-conserving approach to incorporate wave-wave nonlinear effects within the framework commonly used to describe wave-particle nonlinearities. In particular, the effects of zonal mode generation on the determination of the saturation amplitude of energetic particle (EP)-driven Alfvénic instabilities is studied. The model assumes that the zonal perturbations grow at a rate twice that of the original (pump) wave, consistent with a beat-driven (or force-driven) generation mechanism. The evolution and saturation of the mode amplitude are investigated both analytically and numerically within our reduced model assumptions, in both the collisionless and scattering‐dominated regimes. These studies underscore the crucial role of sources and sinks in capturing the impact and the role of beat‐driven zonal perturbations on mode evolution. In the realistic case of saturation set by sources and sinks, we discuss the role of a finite amplitude zonal mode in reducing microturbulent particle scattering, thus limiting the energy source for the resonant mode. We then discuss comparisons between the model’s predictions and simulation results. 
The model reproduces key features observed in gyrokinetic simulations as the reduction in saturated mode amplitude and the onset of wave–wave nonlinear effects as functions of mode growth rate and amplitude. Thanks to its simplicity, it can be readily implemented into codes based on reduced models, thereby improving their predictive capability for strongly driven instabilities.
\end{abstract}

\maketitle
\ioptwocol

\section{Introduction}\label{intro}

Fast ions, produced either by auxiliary heating or fusion reactions, can drive Alfv\'enic instabilities that may degrade plasma confinement and ultimately limit fusion performance  \cite{Salewski_2025}. When these instabilities reach significant amplitudes, they can drive transport of energetic particles (EPs), reducing their effectiveness in transferring energy to the bulk plasma for heating. In addition, this transport may lead to EP losses, which can be highly localized and, in some cases, pose a serious threat to the first wall of a fusion device. In this context, understanding the saturation of EP-driven instabilities is a critical issue in magnetically confined fusion plasmas, as it is directly connected to the level of EP transport that they induce. Over the past decades, extensive research has focused on understanding and predicting the saturation amplitude of these instabilities, which is essential for both theoretical studies and practical applications for reactor design. 

A significant body of work has been dedicated to exploring the role of wave–particle interactions in the evolution of these instabilities \cite{Breizman_2011}. In this context, the Berk–Breizman model (for representative work see \cite{BB1990_1,Berk_1996,Berk_1997_plA}) serves as a cornerstone for interpreting how the free energy available in the EP distribution function is gradually depleted through resonant interactions with the unstable mode. This approach provides a quantitative framework for describing the nonlinear dynamics of wave–particle interactions, especially under conditions where sources and sinks contribute to the overall dynamics of the EP distribution. Here, the sources can be generally associated with EP scattering processes, while the sinks correspond to the mode background damping rates.
In the presence of enough stochasticity, the full nonlinear problem can be reduced \cite{Duarte_2019} to a quasilinear (QL) formulation \cite{Kaufman_1972,Berk_1995_NF}, where a diffusive transport governs the resonant particle dynamics.
As a numerical implementation of the wave–particle QL framework, the Resonance Broadened Quasilinear (RBQ) framework extends these ideas by incorporating geometry effects and accounting for pitch angle scattering resonance broadening. The RBQ model has proven to be numerically effective in predicting the saturation amplitudes of multiple EP-driven modes in realistic tokamak scenarios in both regimes of isolated and overlapping resonances \cite{GorelenkovPoP19}. 

Despite the success of wave–particle interaction models, theory and simulation studies have shown that wave–wave nonlinearities can also play a crucial role in limiting the growth of EP-driven instabilities when they reach a large enough amplitude \cite{Todo_2010,ChenY_2018}. Among these nonlinear effects, the generation of zonal modes (ZM), characterized by zero frequency and toroidal and poloidal mode numbers $n = m = 0$ has emerged as particularly relevant for influencing mode saturation. These modes have attracted growing attention in recent works also due to their potential role in suppressing core thermal plasma turbulence \cite{Mazzi_2022,Garcia_2024,Du_prl}. On this note, we emphasize that the complex system involving EPs, EP-driven instabilities, ZMs, and turbulence has been recently subject of numerous studies. While EP-driven modes are typically associated with EP losses and confinement degradation, they can also interact, both directly and indirectly, with turbulence and turbulence-driven ZMs in ways highlighted by recent works \cite{Wilkie_2018,DiSiena_2018,DiSiena_2019,liu_2022,FIRE_2022,Hahm_2023,Citrin_2023,Ruiz_2025,Na_2025}. Although a conclusive picture has not yet emerged, these results open the door to scenarios in which, despite EPs destabilizing modes that degrade their own confinement, their effect on turbulence could nevertheless lead to an overall improvement in plasma performance.

The theoretical framework describing how ZMs can be nonlinearly excited by fast-ion-driven instabilities, and their subsequent impact on mode saturation, has been extensively developed starting with the seminal works by Chen and Zonca \cite{Chen_Zonca_2012,Chen_Zonca2016}.
Nonideal, compressible or nonuniformity/geometric effects can lead to the so-called “breaking of the pure Alfv\'enic state” \cite{Chen_2013}, i.e., the breaking of the exact balance between Reynolds and Maxwell stresses. This facilitates self-coupling and modulational instabilities that, in turn, can lead to the generation of ZM. 
Two primary mechanisms have been identified for the nonlinear generation of ZMs:
\begin{itemize}
    \item \textit{Modulational instability (spontaneous generation):} Similar to what has been established for drift wave turbulence \cite{Chen_2000,Diamond_2005}, ZMs can be nonlinearly generated by Alfv\'en modes via modulational instability mechanisms \cite{Chen_Zonca_2012,Chen_Zonca2016}. Often referred to as spontaneous generation, this mechanism requires the amplitude of the fast-ion-driven mode to exceed a certain threshold and predicts a ZM growth rate proportional to  fast-ion-driven mode amplitude.
    \item \textit{Beat-driven (force-driven) generation:} Commonly observed in numerical simulations \cite{Todo_2010,ChenY_2018,Biancalani_2020,Spong_2021,Brochard_2024,Sama_2024,Chen_2025} and described theoretically in \cite{Qiu_2017,Qiu_2023,Chen_2025}, this mechanism involves the generation of a zero-frequency $(n, m) = (0, 0)$ ZM through the nonlinear beating of a finite-frequency, finite-$(n, m)$ mode with itself. Unlike the modulational instability, this process is thresholdless and leads to a ZM growth rate twice that of the fast-ion-driven mode.
\end{itemize}
In this work, we focus on the beat-driven generation of ZMs, as it is the mechanism most commonly observed in simulations and does not require a threshold to be triggered. Our analysis will thus concentrate on the linear and early nonlinear stages of the mode evolution, where this mechanism is most relevant. It is important to note that additional nonlinear effects, such as the spontaneous generation of ZMs and nonlinear transport processes on long time scales \cite{Falessi_2023}, may become significant at later stages and could ultimately influence the final saturation amplitude. Our model aims to provide a first attempt to include wave-wave nonlinearities in the perturbative wave-particle framework, to determine a first estimate of the wave-wave nonlinearities effect on the mode amplitudes, thus improving the predictive capabilities of the model.

The challenge, however, lies in developing a unified theoretical framework that can capture both wave-particle and wave-wave mechanisms concurrently. An accurate prediction of the mode saturation for conditions that would lead to large amplitudes requires a treatment that self-consistently integrates the contributions from both resonant particle dynamics and the nonlinear generation of ZMs. Our approach builds on the established foundations of wave–particle interaction perturbative theory while introducing an intuitive mechanism to account for the nonlinear wave-wave interactions.
By comparing the predictions of our model with available simulation results, we aim to provide insights into the relative importance of these coexisting nonlinearities and their implications for the operational regimes of tokamak experiments. 
Our results will highlight the crucial role of sources and sinks for the pump mode resonant interaction to correctly capture the effects of beat-driven ZM on saturation. 

The paper is organized as follows. In Section \ref{1Dmodel}, within the same formalism used to describe the wave-particle nonlinear behaviour, we introduce our reduced model to account for the beat-driven generation of ZMs. The extended model, consisting of coupled equations for pump and ZM amplitudes, is then studied to determine saturation amplitudes for limiting analytic cases in Sections \ref{collisionless} and \ref{collisional} for collisionless and collisional limits respectively. In Section \ref{BOT}, the analytic results are compared with simulations using the BOT code \cite{Lillley_2010}, which has been extended to solve the coupled equation system presented in the previous sections. Finally, conclusions are reported in Section \ref{Conclusions}.

\section{Reduced model for mode amplitude evolution}\label{1Dmodel}
We analyze the evolution of a single eigenmode destabilized by wave–particle resonant interactions under the assumption of ideal MHD fluctuations, for which the parallel electric field vanishes, $\delta E_\parallel = 0$. Our focus is on shear Alfv\'en-type instabilities driven by energetic ions. For these modes, the parallel magnetic perturbation can be assumed to be negligible, $\delta B_\parallel \approx 0$, consistent with wave propagation along the equilibrium magnetic field direction $\boldsymbol{b_0} = \boldsymbol{B_0}/|\boldsymbol{B_0}|$. We focus on a single mode, referred to as the \textbf{pump} wave, which is assumed to oscillate in time as $e^{-i\omega_p t}$, with real frequency $\omega_p$.
Given the wavevector $\boldsymbol{k}$, the dispersion relation is $\omega_p = v_A k_\parallel$, where $k_{\parallel} = \boldsymbol{k} \cdot \boldsymbol{b_0}$ and $v_A = B/\sqrt{\mu_0\rho_i}$ is the Alfv\'en speed in a plasma with ion mass density $\rho_i$.
For simplicity, we consider a mode with single toroidal and poloidal mode numbers $n$ and $m$, respectively. The electrostatic potential is modeled as
\begin{equation}
    \phi_p(\boldsymbol{x},t) = \hat{\phi}_p(t) \, \alpha(r) \, e^{im\theta + in\varphi} \, e^{-i\omega_p t} + \text{c.c.}
\end{equation}
The slowly varying radial envelope is split into a time-dependent amplitude $\hat{\phi}_p(t)$ and a fixed radial mode structure $\alpha(r)$. We also assume that the envelope amplitude evolves on a slower timescale than the oscillatory part of the wave, $\omega_p\gg|\partial_t\hat{\phi}_p(t)|$.

Finally, we introduce the perpendicular electric field of the pump wave associated with the electrostatic potential $\phi_p$ as $ \boldsymbol{E}_p = -\nabla_\perp \phi_p$.
Neglecting the slow variation of the radial envelope, the magnitude of the electric field can be approximated as $|\boldsymbol{E}_p| \approx | k_\perp \hat{\phi}_p(t) \alpha(r)|$.
We can now express the mode energy density in terms of the electrostatic potential envelope. The total mode energy density is the sum of kinetic and magnetic contributions:
\begin{equation}
    w = \frac{1}{2} \rho_i |\boldsymbol{u}_p|^2 + \frac{|\delta \boldsymbol{B}_p|^2}{2\mu_0},
\end{equation}
where $\boldsymbol{u}_p = (\boldsymbol{E}_p \times \boldsymbol{B_0}) / B_0^2$ is the $\boldsymbol{E} \times \boldsymbol{B}$ fluid velocity. For shear Alfv\'en waves, the kinetic and magnetic energy contributions are approximately equal \cite{Priest1982}. Therefore, the total mode energy can be written as
\begin{equation}
    \mathcal{E}_p = \int d \mathcal{V} \, \frac{\rho_i}{B_0^2} \, k_\perp^2 \, |\hat{\phi}_p(t) \alpha(r)|^2.
\end{equation}
with volume element $ d \mathcal{V}$.
We define the squared time-dependent mode amplitude, which is proportional to the wave energy, as 
\begin{equation}
    A_p^2(t) =\, \hat{\phi}_p^2(t) \int  d \mathcal{V}  \, k_\perp^2  |\alpha(r)|^2.\label{amplitudeP}
\end{equation}

We study the dynamics of $A_p^2(t)$ affected by the self-interaction of the pump mode, leading to the nonlinear generation of a zero-frequency perturbation with toroidal and poloidal mode numbers $n = m = 0$. The drive of zonal structures is accounted as an effective dissipation mechanisms, introducing the additional damping term $\gamma_{d,ww}(t)$. It represents the energy transfer from the pump wave to the ZM and accounts for this wave-wave nonlinear effect. The modified evolution equation becomes:
\begin{equation}
    \frac{dA_p^2(t)}{dt} = 2\left[\gamma_{NL}(t) - \gamma_{d,p} - \gamma_{d,ww}(t) \right] A_p^2(t),
    \label{eq:amplitudeP1}
\end{equation}
where the total growth rate has been decomposed into the nonlinear growth rate $\gamma_{NL}(t)$, arising from wave–particle interactions (depending on the mode frequency, structure, and the evolving distribution function of the energetic particles), a constant background damping rate $\gamma_{d,p}$ and the additional damping term $\gamma_{d,ww}$ from nonlinear wave generation.
Effectively, Eq.~\eqref{eq:amplitudeP1} with $\gamma_{d,ww}(t)=0$ is the amplitude evolution equation used in the wave–particle interaction framework. 

An additional coupled equation is required to describe the evolution of the ZM amplitude:
\begin{equation}
    \frac{dA_z^2(t)}{dt} = 2\left[\gamma_{z}(t) - \gamma_{d,z} \right] A_z^2(t),
    \label{eq:amplitudeZ1}
\end{equation}
where $\gamma_{z}(t)$ represents the growth rate of the ZM due to the nonlinear drive from the pump wave, and $\gamma_{d,z}$ is the damping rate of the zonal mode to the background plasma. The ZM amplitude is defined in the same way as for the pump mode starting from the beat-driven ZM energy that, contrary to the case of the Alfv\'enic perturbation, we consider to be stored mostly as fluid kinetic energy of the poloidal flow:
\begin{equation}
    \mathcal{E}_z = \int  d \mathcal{V}  \, \frac{\rho_i}{ 2B_0^2} \, k_z^2 \, |\hat{\phi}_z(t) \alpha_z(r)|^2.
\end{equation}
where we introduced the ZM electrostatic potential amplitude $\hat{\phi}_z(t)$ with radial structure $\alpha_z(r)$ and radial mode number $k_z$. The amplitude is then defined as:
\begin{equation}
    A_z^2(t) =  \, \hat{\phi}_z^2(t) \int  d \mathcal{V}  \, k_z^2|\alpha_z(r)|^2.\label{amplitudeZ}
\end{equation}

To couple the evolution equations of the pump wave and the ZM, we impose energy conservation for the beat-driven generation process, thereby relating the nonlinear damping terms in Eq.~\eqref{eq:amplitudeP1} to the corresponding growth terms in the zonal equation.
The power lost from the AE due to beat-driven nonlinear generation is:
\begin{equation}
    \mathcal{P}_{p\rightarrow ww} = \gamma_{d,ww}(t)\mathcal{E}_p(t)
\end{equation}
and must be balanced by the power gained by the ZM $ \mathcal{P}_{p\rightarrow ww} = \mathcal{P}_{z} $
% We will focus on the ZM and assume that the power flowing to the zonal component is the comparable to the one towards the harmonic as $\mathcal{P}_{z} \approx \mathcal{P}_{hh}$. 
from which one obtains
\begin{equation}
   \gamma_{d,ww}(t) A_p^2(t) = \gamma_{z}(t) A_z^2(t)/2, \label{power-balance}
\end{equation}
from which it is possible to rewrite the wave-wave damping of the pump wave in Eq.~(\ref{eq:amplitudeP1}) in terms of the ZM growth rate as $\gamma_{d,ww(t)} = \gamma_{z}(t)A_z^2(t)/2A_p^2(t)$.

As discussed in Sec. \ref{intro}, numerous analytical and numerical studies have described possible mechanisms for the generation of ZM from nonlinear wave-wave interactions. An important remark is needed at this point. Here, we do not aim to add to this part of the literature, and to describe the specific mechanism by which the ZM is generated by the pump wave. Instead, we seek to leverage the extensive body of research on these processes to, at least in a reduced way, describe and understand how the nonlinear generation of ZM can influence the saturation amplitude of the pump. 
To do so, we focus on the mechanism that appears to dominate the early stages of the simulations and is likely major in limiting the pump amplitude: the beat-driven generation of ZM.
This process is threshold less and leads to ZM growth following  $\gamma_z = 2\left[\gamma_{NL}(t) - \gamma_{d,p} - \gamma_{d,ww}(t) \right]$ where $\left[\gamma_{NL}(t) - \gamma_{d,p} - \gamma_{d,ww}(t) \right]$ represents the instantaneous total growth rate of the pump wave. This mechanism serves as the primary nonlinear drive for ZM during the linear and early nonlinear phases of pump evolution. Other nonlinear mechanisms, such as the modulational instability \cite{Chen_Zonca_2012}, can be important in later stages if the pump wave amplitude becomes large enough. In such cases, these other effects may significantly impact the mode saturation after the beat-driven generation ceases when the pump wave stops growing and $\left[\gamma_{NL}(t) - \gamma_{d,p} - \gamma_{d,ww}(t) \right] = 0$. 
Only when neglecting other nonlinear ZM effects, if the pump wave exhibits no overshoot oscillations, the amplitude at the time for which $\left[\gamma_{NL}(t) - \gamma_{d,p} - \gamma_{d,ww}(t) \right] = 0$ represents the saturated pump amplitude. In more general cases the pump amplitude can, however, continue to evolve significantly after its first peak, which does not necessarily correspond to the amplitude at saturation.

In the following, we will assume $\gamma_z = 2\left[\gamma_{NL}(t) - \gamma_{d,p} - \gamma_{d,ww}(t) \right]$. While this limits the scope of the model to capturing the influence of the ZM on the pump wave due to the beat-driven generation, it allows for simplification of the model. The resulting amplitudes can be considered as a zeroth-order estimate of the effect of wave-wave nonlinearities on AE evolution. While additional nonlinear mechanisms may become relevant at later stages, their impact will build upon the amplitudes established by this initial approximation. 
Within this assumption, using the relation of Eq.~\ref{power-balance} we obtain the ZM growth rate from $\gamma_{z} = 2\left[\gamma_{NL}(t) - \gamma_{d,p} - \gamma_{z}(t)A_z^2(t)/2A_p^2(t)\right]$:
\begin{align}
\gamma_z = 2\frac{\left[\gamma_{NL}(t) - \gamma_{d,p} \right]}{1+A_z^2(t)/A_p^2(t)}
    \label{forcedriven}
\end{align}
Substituting relation (\ref{forcedriven}) in the coupled equations (\ref{eq:amplitudeP1},\ref{eq:amplitudeZ1}):
\begin{align}
    &\frac{dA_p^2(t)}{dt} = 2\frac{\left[\gamma_{NL}(t) - \gamma_{d,p} \right]}{1+A_z^2(t)/A_p^2(t)} A_p^2(t),
    \label{eq:amplitudeP3}
    \\
    &\frac{dA_z^2(t)}{dt} = 2\left\{2\frac{\left[\gamma_{NL}(t) - \gamma_{d,p} \right]}{1+A_z^2(t)/A_p^2(t)} - \gamma_{d,z} \right\} A_z^2(t).
    \label{eq:amplitudeZ3}
\end{align}
From these coupled system of equation it becomes apparent how the generation of a ZM via beat-driven process will reduce the net growth rate of the pump wave through the denominator $1+A_z^2(t)/A_p^2(t)$. Moreover, if the background damping rate of the ZM, $\gamma_{d,z}$, can be assumed to be negligible with respect to the nonlinear drive, the two equations are formally the same, leading to the condition valid at any time
\begin{equation}\label{forceK}
    A_z(t)= K_{z}A^2_p(t) 
\end{equation}
where $K_{z}$ is a parameter that determines the initial condition $ A_z(0)= K_{z}A^2_p(0)$. 
In the presence of a background ZM damping rate $\gamma_{d,z}$, a similar expression relates $A_z(t)$ and $A_p(t)$: $A_z(t) = K_{z} A_p^2(t)\, e^{-\gamma_{d,z} t}$ describing how the beat-driven relation in Eq.~(\ref{forceK}) is lost over time due to the damping of the ZM. For the remainder of this work we will, however, neglect $\gamma_{d,z}$ for simplicity.

While we will consider $K_{z}$ as a free parameter of our model, it depends on the pump wave characteristics and is essentially linked with the self-beating of the wave and with the breaking of the pure Alfv\'enic state. The existing analytic theory describing the beat-driven process \cite{Chen_Zonca2016,Qiu_2023,Chen_2025} can be leveraged to obtain a approximate first estimate of the parameter $K_{z}$. From the latest work on the subject, we consider Eq.~(31) of Ref.~\cite{Chen_2025}: $\phi_z \approx [c(1 + \eta_i)/B_0\omega_p^2] \partial_r (k_{\theta} \omega_{*i} |\phi_p|^2)$, in CGS units, with $\eta_i = |\nabla \log T_i|/|\nabla \log n_i|$ and $\omega_{*i} = (cT_i/eB_0)(\boldsymbol{k}\times\boldsymbol{b_0})\cdot \nabla \ln{n_i}$. It is worth noting here that the zonal potential and the squared pump potential are related by the nonuniformity of the background plasma, which is determining the radial gradient of the mode structure, and which is one of the mechanisms that can lead to the breaking of the pure Alfv\'enic state. 
From Eqs.~(\ref{amplitudeP}), (\ref{amplitudeZ}) and relation (\ref{forceK}), we get 
\begin{equation}\label{kzestimate}
K_{z} \approx \frac{ (\int  d \mathcal{V}  \, k_z^2|\alpha_z(r)|^2 )^{1/2}}{ \int  d \mathcal{V}  \, k_\perp^2|\alpha(r)|^2} \frac{c(1 + \eta_i)}{B_0\omega_p^2}  k_r k_{\theta} \omega_{*i}, 
\end{equation}
where we considered $\partial_r \sim k_r$.
For a first estimate, we consider the Alfv\'enic mode in the DIII-D tokamak described in Ref.~\cite{ChenY_2018}, where also the effects of self-generated ZM is discussed. The $n=4$ RSAE mode under study has frequency $\omega_p\approx 4.5 \: 10^5 \text{rad/s}$. From Figure 3 of Ref.\cite{ChenY_2018} the dominant poloidal mode number is $m=12$ and the mode is located at $r\approx 30  \text{cm}$ with radial extent $\Delta r \approx 10 \text{cm}$. As a simplified estimate we take $k_r\approx 1/\Delta r = 0.15 \text{cm}^{-1}$ and $k_\theta\approx m/r = 0.4 \text{cm}^{-1}$. We consider the quantity $(\int  d \mathcal{V}  \, k_z^2|\alpha_z(r)|^2)^{1/2}/\int  d \mathcal{V}  \, k_\perp^2|\alpha(r)|^2 $ of order unity. The field on axis is $B_0\approx 2.2 T$, the ion temperature at the mode location is $T_i\approx 1 keV$, and from Figure 1 of Ref.\cite{ChenY_2018} we estimate $\nabla \ln{n_i} = (15 \text{cm})^{-1}$, $\eta_i\approx 3$ and $\omega_{*i} \approx 30 \text{kHz}$ we obtain $K_{z} \approx 2 \text{statV}^{-1}$.

To solve the coupled Eqs.~(\ref{eq:amplitudeP3}) and (\ref{eq:amplitudeZ3}) and describe the evolution and saturation of the EP-driven instability, one must solve the kinetic problem to determine $\gamma_{NL}$. However, even in the absence of self-generated ZMs, analytic progress is limited. Theoretical frameworks have been developed\cite{BB1990_1,BB1990_2,Berk_1990_3,Berk_1996,BerkPPR1997,Wong_1997} to estimate the saturation amplitude. From those works, approximate analytic expressions for $\gamma_{NL}$ have been derived \cite{Duarte_2019NF,Lestz_2021,Devin_inprep}. These results are only valid within specific asymptotic limits, and more general cases require numerical investigation. 

In the following, we will explore analytically tractable cases that can provide insight into the primary mechanisms responsible for the observed reduction in the pump amplitude in the presence of self-generated ZMs. Building on this analytical understanding, we will expand our analysis using the one-dimensional nonlinear kinetic BOT code \cite{Lillley_2010} to numerically compute $\gamma_{NL}$ and compare those results with our model predictions.

\section{Analytical collisionless saturation}\label{collisionless}
The first limit we study is the collisionless saturation of the pump wave, described in terms of wave-particle trapping \cite{Fried_1971,Dewar_1973}. In this scenario, the saturation mechanism consists in the mode growth until the phase mixing of resonant particles produces a local plateau in the resonance region of the distribution function, as proposed in the context of collisionless damping in \cite{Mazitov_1965,ONeil_1965}. This process effectively depletes the free energy available to drive the resonant mode, leading to its saturation. 

As briefly discussed in Ref.~\cite{Berk_2012}, the saturated state of the system is influenced by the dynamical path taken to reach it. In this context, $\gamma_{NL}(t)$ becomes an integral functional of the entire time history and is generally not known in closed form.

On the other hand, following Ref.~\cite{Berk_1995_NF}, the mode's saturation amplitude can be derived using kinetic and wave momentum conservation. During the nonlinear evolution, the resonant particles lose their initial momentum to the wave as the distribution function is flattened. Saturation occurs when the slope of the resonant particle distribution function at the resonance is reduced to zero, indicating that all the available momentum has been transferred to the wave.
In the absence of self-generated ZMs, external sources, and sinks, the mode is expected to saturate due to wave-particle trapping, satisfying the well-known relation $\omega_b \approx 3.2\,\gamma_L$ \cite{Fried_1971}. The particle bounce (or trapping) frequency of deeply trapped particles within the phase space island due to the resonance, is given by \cite{Duarte_2017_PoP}:
\begin{equation}
    \omega_b = \left[ 2A_p(t)\, V_{n,l}(I_r) \left. \frac{\partial \Omega_{n,l}}{\partial I} \right|_{I=I_r} \right]^{1/2},
\end{equation}
where $V_{n,l}$ are the matrix elements of the wave-particle interaction, and $l$ is an integer. The resonant condition is $\Omega_{n,l}(I) = n\omega_\phi - l\omega_\theta = \omega_p$, where $I = -P_\phi/n|_{\mathcal{E}'}$ is the relevant variable describing the dynamics at the resonance (in terms of the invariants of unperturbed motion - particle energy $\mathcal{E}$ and canonical toroidal momentum $P_\phi$ - at constant $\mathcal{E}' = \mathcal{E} + \omega_p P_\phi/n$) and $I_r$ is its value at the resonance. The linear growth rate of the pump wave is defined as
\begin{equation}
    \gamma_L = \frac{\omega_p \pi}{\Lambda} \sum_l \int d\Gamma\, |V_{n,l}|^2\, \delta(\Omega_{n,l} - \omega_p)\, \frac{\partial F}{\partial I},
    \label{eq:gammaL}
\end{equation}
where $\delta(x)$ is a Dirac delta function, $\Lambda$ is the mode inertial energy over amplitude squared \cite{Duarte_2019NF}, the summation is over the $l$ integers, the integration is over the phase space with $d\Gamma$ denoting the phase-space element and $F$ is the equilibrium particle distribution function in the absence of the mode.
For our analysis, it is important to note that $\omega_b \propto A_p^{1/2}$. This proportionality is also captured in the simplified 1D bump-on-tail instability model, where 
\begin{equation}\label{1DomegaB}
    \omega_b = \sqrt{q_{EP} A_{p,1D} k / m_{EP}},
\end{equation} with $q_{EP}$ and $m_{EP}$ being the charge and mass of the energetic particles, and $A_{p,1D}$ and $k$ the electric field amplitude and wave number of the electrostatic mode.

In order to study the effect of the beat-driven ZMs, we assume that in our model the only effect of the wave-wave nonlinearity is the flow of energy from the pump to the self-generation of the ZMs. The beat-driven generation does not impact the resonance condition via Doppler shift or the equilibrium and the presence of a finite amplitude ZM does not affect the wave-particle trapping. These assumptions may be overly restrictive in the presence of strong wave–wave nonlinearities, but they allow us to consider that the same amount of free energy is available in the distribution function with and without nonlinear wave generation, and that this energy depends solely on the pump wave. Since the presence of the beat-driven generation does not affect the free energy available in the system, it is possible to analyze a simplified toy model of the problem based on the one-dimensional bump-on-tail instability, extending the analysis of Refs.~\cite{Berk_1995_NF,White_2019} based on kinetic and wave momentum conservation. 

In contrast to Ref.\cite{Berk_1995_NF}, in presence of nonlinear wave-wave generation, the wave momentum at saturation is the combination of pump and nonlinearly generated waves. Within the one-dimensional model, it is not appropriate to consider the momentum gained by the ZM, since we are neglecting all the information related to the radial mode structure essential for an appropriate description of the ZM. However, some qualitative understanding can still be gained by considering the beat-driven generation of a higher harmonic wave with twice the frequency and wave number. Although in realistic scenarios this higher harmonic is typically subject to strong dissipation \cite{Todo_2012} and remains at very low amplitude compared to the beat-driven zonal counterpart, in the collisionless regime, where damping terms are absent, it can play an equivalent role to that of the ZM. Specifically, it provides an additional degree of freedom where the energy extracted by the pump from the particles can be redirected, thereby limiting the pump’s growth. The same arguments that we used to derive the coupled Eqs.~(\ref{eq:amplitudeP3}) and (\ref{eq:amplitudeZ3}) can be used for the higher harmonic beat-driven amplitude $A_{hh}(t)$. In the absence of sources and sinks a relation $A_{hh}(t) = K_{hh} A_{p,1D}^2(t)$ is then valid at any time.

Following the same procedure as in Ref.\cite{Berk_1995_NF,White_2019}, the kinetic momentum available for the waves from the gradient flattening is $\Delta W_{kin} \propto \omega_b^{3}$. In terms of the wave amplitude one can write $\Delta W_{kin} = c_{L} A_{p,1D}^{3/2}$ with $c_{L}$ being a coefficient that depends on the linear growth rate of the pump. In the absence of beat-driven generation, the kinetic momentum is transferred only to the pump wave. At the saturation, the kinetic momentum $\Delta W_{kin}$ lost from the distribution due to the flattening process has been gained by the pump wave momentum, which is proportional to the amplitude squared: $W_{p,sat} = kA_{p,1D}^2/\omega_p$. The momentum balance $W_{p,sat} = \Delta W_{kin}$ gives 
\begin{equation}
    \frac{k}{\omega_p}A_{p,1D}^2 =  c_{L} A_{p,1D}^{3/2},
\end{equation}
from which, using the definition of Eq.~(\ref{1DomegaB}):
\begin{equation}
    \omega_b =\sqrt{\frac{q_{EP}}{m_{EP}k}}\omega_p c_L
\end{equation}
Knowing that for the collisionless saturation in absence of nonlinear wave generation $\omega_b \approx 3.2\gamma_L$ holds \cite{Fried_1971,Dewar_1973,Berk_1995_NF}, we can determine the coefficient $c_{L}$ such that $\omega_b c_L \sqrt{q_{EP}/m_{EP}k} \approx 3.2\gamma_L$ to reproduce the expected saturation from numerical results.

In the presence of nonlinear wave generation, the total momentum balance requires $W_{p,sat} + W_{hh,sat} = \Delta W_{kin}$, where $W_{hh,sat}$ is the momentum of the beat-driven wave at saturation. Within our assumption, the kinetic momentum that the mode can extract can be still written as $\Delta W_{kin} = c_{L} \omega_b^3$. From the relation $A_{hh}(t) = K_{hh} A_{p,1D}^2(t)$ one can write the higher harmonic wave momentum as $W_{hh}(t) = 2kK_{hh}^2A_{p,1D}^4/2\omega_p$. The momentum balance reads:
\begin{align}
     &W_{hh,sat} + W_{p,sat} = \Delta W_{kin}\\
     &\frac{k}{\omega_p}K_{hh}^2A_{p,1D}^4 + \frac{k}{\omega_p}A_{p,1D}^2 =  c_{L} A_{p,1D}^{3/2} 
\end{align}
Defining $x = A_{p,1D}^{1/2}$, the relation for the pump at saturation follows:
\begin{equation}
     K_{hh}^2x^5 + x = 3.2\gamma_L\sqrt{\frac{m_{EP}}{kq_{EP}}}  
    \label{quintic2}
\end{equation}
which can be solved numerically. Equation (\ref{quintic2}) shows that in case of no self-generation the standard solution is recovered, i.e. with $K_{hh} = 0$ we obtain the expected $\omega_{b,sat}\propto\gamma_L$. With beat-driven generation, $ K_{hh}\neq0$, the energy available to the pump is reduced for two main reasons. Firstly, part of the kinetic energy available in the distribution is "rerouted" to the beat-driven mode. Secondly, due to this leak of energy, the pump wave growth is reduced, and its amplitude at each time is smaller compared to the case with $ K_{hh} = 0$. Therefore, the total free energy that the mode can extract from the distribution is also reduced with respect to the $ K_{hh} = 0$ case. The additional energy lost due to the nonlinear wave-wave interactions, as any additional damping of the pump wave, could enhance hole and clumps formation and lead to chirping behaviour \cite{BIERWAGE2021}. In our perturbative analysis we however focus on cases characterized by a monotonic growth followed by steady saturation.

\begin{figure}
    \centering
    \includegraphics[width=0.99\linewidth]{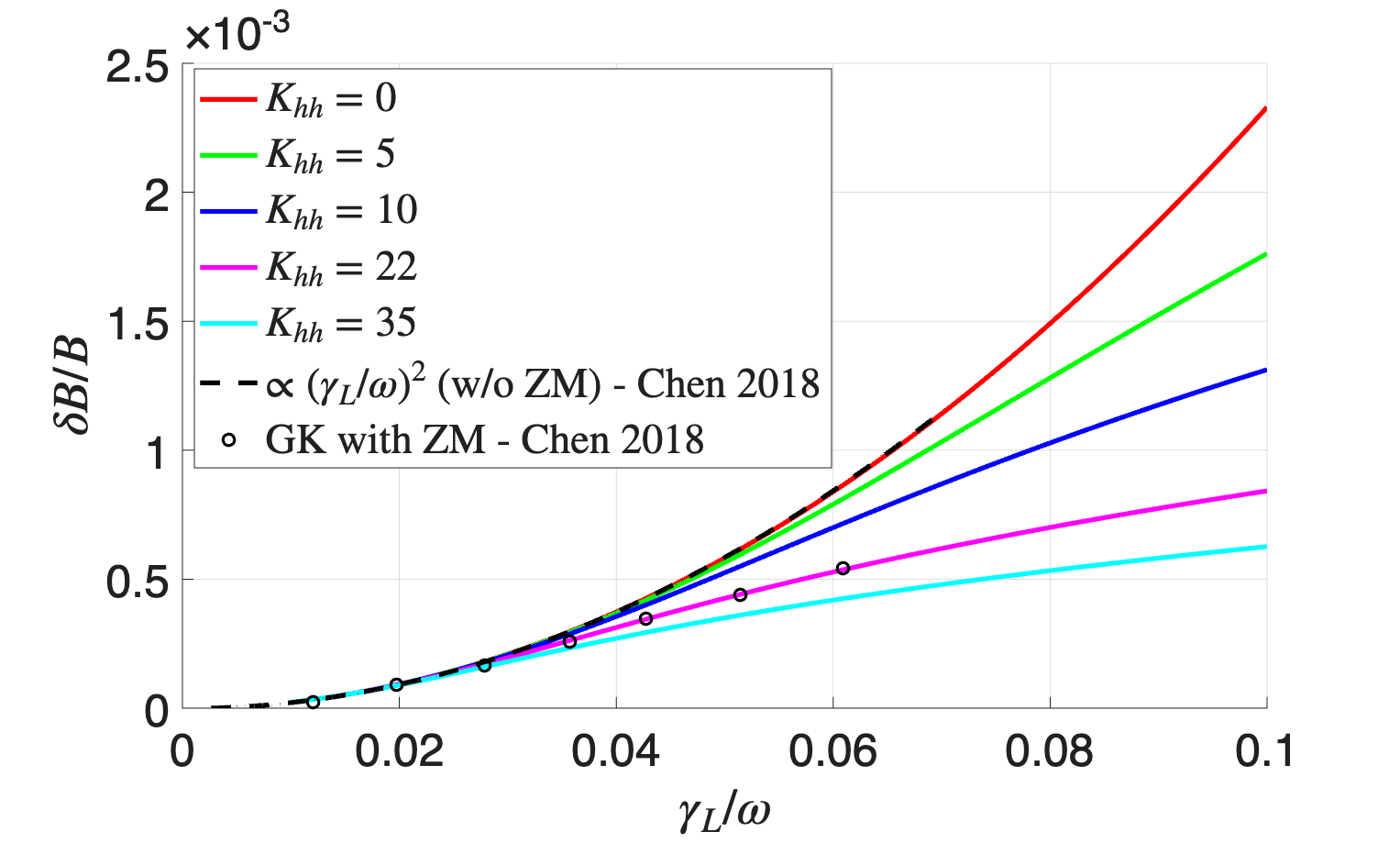}
    \caption{Value of $\delta B_r/B$ at saturation for different values of the model parameter $K_{hh}$. The normalized model result are compared with the data published in Chen 2018 \cite{ChenY_2018} coming from nonlinear collisionless gyrokinetic simulations using the code GEM. The non-zonal curve from \cite{ChenY_2018} follows $\delta B_r/B = 0.233(\gamma_L/\omega_p)^2$ and is reproduced as expected by the $ K_{hh}=0$ case. Good agreement between our model and the GEM simulations is found for $ K_{hh}=22$.}
    \label{fig:comparison}
\end{figure}

While the toy model based on momentum balance for the 1D bump-on-tail instability can be extended to include the beat-driven higher harmonic instead of ZMs, we propose that the same mechanisms can qualitatively describe how the ZM reduces the saturation amplitude of the pump mode in more realistic scenarios.
Although a quantitative comparison with more complete models would require more investigation, the trend observed here compares favorably with collisionless simulations that showed the discrepancy in saturation with and without ZM generation increasing with increasing linear growth rate. In Ref.\cite{ChenY_2018} the threshold in linear growth rate for the thermal plasma nonlinearities to play a role was identified as $\gamma_L/\omega_p \approx 3\%$. In Fig.~\ref{fig:comparison} we report Chen's 2018 \cite{ChenY_2018} GEM results compared with our model for different values of the $K_{hh}$ parameter. The value $ K_{hh} \approx 22$ reproduces the GEM results in presence of self-generation of ZM. 

While not rigorous, a first attempt to estimate the effect of beat-driven ZMs on the pump wave saturation amplitude in the collisionless case can be obtained by evaluating the constant $K_{z}$ from Eq.~\ref{kzestimate} and solving the quintic equation using $K_{hh} \sim K_{z}$. 

As an example, our estimate of the parameter $K_{z}$ for the case of Ref.\cite{ChenY_2018}, obtained from Eq.~(\ref{kzestimate}), gives $K_{z}\sim 2$, about an order of magnitude smaller than the value required for the simple 1D model to reproduce the simulation results. Such an underestimation is to be expected for the reduced model, since the analytic estimate accounts only for a subset of the nonlinear dynamics. In contrast, the simulations of Ref.~\cite{ChenY_2018} capture both the ZM ($n=0$) and the higher harmonic ($n=8$) generations, each influencing the evolution of the pump mode ($n=4$), along with additional nonlinear effects beyond the beat-driven ZM mechanism.

In Section \ref{BOT}, the analytic model predictions are compared with nonlinear simulations of the coupled equations (\ref{eq:amplitudeP3}-\ref{eq:amplitudeZ3}) using the nonlinear code BOT \cite{Lillley_2010}.

\section{Analytical saturation with sources and sinks}\label{collisional}
In the previous section, we analyzed the case where saturation is achieved through wave–particle trapping in the absence of sources and sinks. However, that scenario is clearly idealized; in realistic conditions, sources and sinks are essential components and must be accounted for when attempting to describe experimental observations.

The source of resonant particles can be effectively modeled by an effective particle scattering, $\nu_{\text{eff}}$, as detailed in \cite{BerkPPR1997,Candy_1999}. To balance the continuous energy injection from such sources, a dissipation mechanism, effectively represented by the finite background damping of the mode, must also be included. The significance of sources and sinks, and thus of particle scattering, in describing resonant wave–particle interactions has been recognized since early studies \cite{zakharov1962,Su_Oberman,BB1990_1}, and has been emphasized more recently in \cite{Callen2014PoPCoulombCollsions,Duarte_2019,Catto_2021,Hamilton_2023,Liu_2024}.

The presence of scattering and dissipation fundamentally distinguishes this system from the collisionless scenario discussed in Sec.~\ref{collisionless}. In the collisionless case, a finite reservoir of free energy is available to drive wave growth, and saturation occurs once this energy is depleted, enabling a description based on conservation arguments. Such approach is no longer valid in the presence of sources. With continuous injection, the available free energy becomes effectively infinite, and without dissipation the mode would grow without bound. It is the dissipation that limits this growth, and the balance between sources and sinks ultimately governs the nonlinear dynamics and saturation level of the instability.

As briefly mentioned in Sec.~\ref{intro}, wave–particle nonlinearities can give rise to a wide range of complex nonlinear behaviour. Depending on the relative strength of sources and sinks, the nonlinear dynamics may evolve toward either steady-state or pulsating regimes.
In general, even in the absence of ZM generation, an analytic expression for $\gamma_{NL}(t)$ is not known. Nonetheless, there exist sufficiently simple limits that allow for analytical progress. For cases where a steady saturated amplitude is expected, saturation estimates near and far from marginal stability can be obtained following Refs.~\cite{BB1990_1,BB1990_2,Berk_1990_3,Berk_2012,Lestz_2021,Devin_inprep}.
A key feature of these estimates is that they are not influenced by the time history of the mode growth. This property becomes particularly important when introducing the effect of ZM self-generation in these regimes. If we analyze Eq.~(\ref{eq:amplitudeP3}) in the presence of sources and sinks, we observe that the self-generation of the ZM enters solely through the ratio $A_z^2(t)/A_p^2(t)$ in the denominator. This term acts to reduce the net growth rate at each instant due to the presence of the ZM. However, the saturation condition, defined by $dA_{p,sat}^2/dt = 0$, is:

\begin{equation}\label{satcondition}
    \gamma_{NL} - \gamma_{d,p} = 0,
\end{equation}
which is not affected explicitly by the extra term due to the ZM generation. 
The reason for this lies in the beat-driven assumption stated by Eq.~(\ref{forcedriven}). The extra damping term introduced by the self-generation of ZM in Eq.~(\ref{eq:amplitudeP1}) vanishes at saturation. Within our assumptions, in the presence of sources and sinks, there is no reduction of the pump saturated amplitude due to the self-generation of ZM alone. This does not mean that the ZM has no effect on the saturated amplitude, but only that the generation by itself does not, in contrast to what we discussed for the collisionless case. 

\subsection{Considerations on the effects of ZM on sources and sinks}\label{sub:sourcessinks}

While there have been numerous studies on the different effects that a finite amplitude ZM could have on the pump wave, a clear comprehensive picture is still missing. The more common picture in the literature describes that the presence of the ZM is expected to reduce the saturated amplitude of the pump with respect to the case with no generation, as we obtained in the collisionless model in Sec.~\ref{collisionless}. 
For the collisional case, the saturation of the pump is determined by the balance of sources and sinks. This means that for the beat-driven ZM to affect the pump saturation it must affect either the source term in $\gamma_{NL}$ or the damping $\gamma_{d,p}$. 
Starting from the background damping of the pump wave, some possibilities have been outlined in the literature, starting from Ref.\cite{Spong_1994}. The presence of the ZM can impact the damping of the pump in mainly two ways: 
\begin{itemize}
    \item The zonal electrostatic potential induces a radial electric field and a subsequent $\boldsymbol{E}\times\boldsymbol{B}$ sheared flow. This flow can break and decorrelate the pump mode structure causing the mode suppression \cite{Spong_2021}.
    \item The zonal magnetic flux and associated zonal currents affect the $q$ profile, causing small perturbations at the mode location. For cases in which the pump is an Alfv\'en eigenmode, these local modifications of the $q$ profile will impact the continuum damping of the mode \cite{Spong_1994}.
\end{itemize}
It has to be noted, however, that a detailed description of how these two mechanisms quantitatively affect the pump damping rate is still lacking.

The other option for the finite amplitude ZM to affect the nonlinear saturation in our model is via modifications of $\gamma_{NL}$. 
Again, different mechanisms that might play a role are possible:
\begin{itemize}
    \item The presence of the ZM might induce a Doppler shift of the resonant condition impacting the wave-particle interaction, thus affecting the nonlinear growth rate \cite{Brochard_2024,Brochard_2025}.
    \item The ZM can interact with the resonant particles altering phase space zonal structures (PSZS) generated from the wave-particle resonance. This interaction can modify the available free energy for the mode to grow, thus affecting $\gamma_{NL}$ \cite{Chen_2025}.
\end{itemize}
Both mechanisms allow for some variability in their impact on $\gamma_{NL}$ as these mechanisms may also lead to an increased pump saturation amplitude.
In Ref. \cite{Brochard_2025}, the ZM-induced Doppler shift is shown to influence the saturation of chirping fishbone oscillations by restricting the regions of the distribution function where the mode can resonate, compared to cases without ZM. Conversely, for a steady-frequency mode, the same Doppler shift would expand the accessible regions of the distribution function, thereby increasing the available free energy and potentially enhancing the mode's growth. A similar argument could apply to the effect of ZM on PSZS. As demonstrated both analytically and numerically in Ref. \cite{Chen_2025}, when the interaction between ZM and resonant EP is isolated from other effects, it can enhance the drive depending on the EP distribution function, leading to an increased saturation amplitude of RSAEs. 
When isolated, neither mechanism appears sufficient to consistently produce to the expected reduction of pump saturated amplitude for a general case of shear Alfv\'en wave driven by fast ions.

While the two mechanisms discussed above describe how $\gamma_{NL}$ can be modified in the presence of a ZM through changes in the wave–particle resonant interaction, an additional possibility is that the ZM affects the particle source itself, namely, the particle scattering. Within the wave–particle nonlinear framework, the effective particle scattering rate $\nu_{\text{eff}}$ represents the source of particles for the resonance. Namely, it restores the gradients and replenishes the free energy in the distribution function. This effective scattering incorporates all mechanisms that can scatter particles in phase space into and out of the resonant island. For a realistic mode with finite radial width, the quantity $\nu_{\text{eff}}$ is an appropriate average over phase space weighted by the mode structure \cite{1999_Gorelenkov}. In the absence of ZM, the expected scaling of the saturation amplitude follows $A_{p,sat}\propto \nu_{\text{eff}}^2$ \cite{petviashvili1999}.
In general, $\nu_{\text{eff}}$ includes contributions from Coulomb pitch-angle collisions as well as anomalous scattering due to turbulence \cite{Lang_2011}.
The latter contribution is particularly relevant for our model since the presence of a finite amplitude ZM can reduce the turbulence level, either through $\boldsymbol{E} \times \boldsymbol{B}$ shearing or by influencing the self-saturation of drift waves. This turbulence suppression in turn reduces anomalous particle scattering and thereby lowers the effective particle source. Although this effect is nontrivial to quantify, its qualitative impact is evident and, unlike the previously discussed mechanisms, it leads to a reduction of the particle source and thus a lower saturation amplitude of the pump wave. 

\subsection{Reduction of particle source due to scattering suppression}\label{sub:subZFsupp}
%\subsection{Time dependent $\nu_{\text{eff}}$ due to ZM shearing}
We proceed analytically by first recalling the results for the nonlinear evolution and saturation amplitude in the absence of ZM generation. Focusing on the regime near marginal stability, i.e., $(\gamma_L - \gamma_{d,p})/\gamma_L \ll 1$, analytic progress becomes feasible as the nonlinear growth rate can be obtained from a cubic equation involving time-delayed integrals \cite{Berk_1996}. 
To simplify the analysis further we consider the additional expanding parameter $\nu_{\text{eff}}/(\gamma_L - \gamma_{d,p}) \gg 1$, consistent with typical experimental conditions in conventional tokamaks \cite{Duarte_2017}. In this regime, as shown in \cite{Duarte_2019NF}, the integro-differential cubic equation becomes effectively local in time. Under these assumptions, the analytic expression for the nonlinear growth rate of the pump wave, in the absence of ZM self-generation, reads \cite{Duarte_2019NF,Lestz_2021}:

\begin{equation}\label{nonlinearg}
    \gamma_{NL}(t) = \gamma_L\left(1-e^{-i\Phi }\frac{\Gamma(1/3)(3/2)^{1/3}}{6}\frac{|\omega_b^2(t)|^2}{\nu_{\text{eff}}^4}\right)
\end{equation}
where $\Phi$ in the term $e^{-i\Phi}$ is a measure of the contribution of the  fast particles to the mode frequency \cite{BerkPPR1997,Breizman_1997,Fasoli_1998}. For Alfv\'en eigenmodes that can be treated perturbatively, this phase should be $\Phi = \pi N$, with $N$ integer. However, for a more general case $\Phi \neq \pi N$ allows for a description with resonant particles affecting the frequency of the mode.
From Eq.~(\ref{nonlinearg}), one can see how the nonlinear growth rate is strongly affected by the source term, i.e. the effective scattering $\nu_{\text{eff}}$.
When these assumptions hold, the evolution of the mode amplitude proceeds through monotonic growth until it reaches a steady saturated state, which satisfies the condition given in Eq.~(\ref{satcondition}). In the absence of ZM self-generation, this condition reduces to:
\begin{align}\label{nozonal_sat}
    \omega_{b,sat} = \sqrt[4]{\frac{6e^{i\Phi }}{\Gamma(1/3)(3/2)^{1/3}}}\left(1-\frac{\gamma_{d,p}}{\gamma_L}\right)^{1/4}\nu_{\text{eff}},
\end{align}
here, although in general $\omega_{b,\text{sat}}$ is a complex quantity, in this work we are only concerned with its amplitude. Accordingly, we take $\omega_{b,\text{sat}}$ to be the absolute value of the saturated bounce frequency.

The first step in incorporating ZM effects is to model its impact on the time-dependent turbulent scattering rate, $\nu_{\text{turb}}(t)$.
To attempt to describe the effect of the ZM on the particle source, we express the effective scattering rate $\nu_{\text{eff}}$ as the combination of effective collisional and turbulent contributions \cite{Duarte_2017,Duarte_2017_PoP}:
\begin{equation}
    \nu_{\text{eff}} = \nu_{\text{coll}} + \nu_{\text{turb}}.
\end{equation}
We then model the influence of self-generated ZM on the particle source, and thus on the pump wave amplitude, through a time-dependent turbulent scattering rate $\nu_{\text{turb}}(t)$. For simplicity, and consistent with conditions in conventional tokamaks \cite{Lang_2011,Duarte_2017_PoP}, we assume that $\nu_{\text{eff}} \ll \nu_{\text{turb}}$, thereby neglecting the collisional contribution and simplifying the notation. To this end, we recall the form of $\nu_{\text{turb}}$ as used in Ref.~\cite{Duarte_2017,Duarte_2017_PoP}, based on the pioneering analysis of Ref.~\cite{Lang_2011}, which describes the EP radial diffusivity due to microturbulence within resonances as:
\begin{equation}
    \nu_{\text{turb}}^3 = D_{EP}\left( \frac{q_{EP}}{m_{EP}}\frac{\partial \psi}{\partial r} \right)^2 \left( \frac{\partial \Omega}{\partial I} \Bigg|_{I_r}\right)^{2},
\end{equation}
where $\psi$ is the flux function.
Here, $D_{EP}$ denotes the EP diffusivity, which is connected to the thermal ion diffusivity $D_{th,i}$ through a proportionality scaling, as discussed for instance in Ref.~\cite{Zhang_2008,Angioni_2009}. Consequently, the effective scattering rate scales as $\nu_{\text{eff}}^3 \propto D_{th,i}$.

To describe the suppression, we assume that the influence of self-generated ZM on turbulence manifests itself with the same $\boldsymbol{E}\times\boldsymbol{B}$ stabilization effect induced by a mean flow \cite{Biglari_1990, Burrell_1997,Terry_2000}. The validity of this assumption corresponds to the situation in which the pump mode exhibits a global structure and the associated nonlinearly generated ZM has a radial scale significantly larger than the characteristic length of turbulence eddies. Under these conditions, analytical progress can be made without introducing an additional coupled equation for drift wave turbulence. The scenario in which the beat-driven ZM directly enters the predator-prey system, used to describe the self-saturation of turbulence via nonlinearly generated zonal flows \cite{Diamond_2005, Itoh_2006, Zhu_Dodin_2020}, is beyond the scope of this article and will be explored in future work. 

Assuming that the relevant turbulence is driven by temperature-gradient instabilities, we consider the phenomenological theory framework of Ref. \cite{Ivanov_2025} to describe the impact of a shear flow on the diffusion parameter $D_{th,i}$.
Due to the assumption of radial scale separation between the ZM and the turbulence, we assume the beat-driven poloidal flow to be seen by the turbulence, at least locally, as 
\begin{equation}
    \boldsymbol{v}_\theta = \gamma_E r \boldsymbol{\hat{\theta}}
\end{equation}
where we approximate a spatial constant shear $\gamma_E$ of the time varying $\boldsymbol{E}\times\boldsymbol{B}$ flow.
For a given pump mode structure, the beat-driven ZM radial structure determines the radial dependence of turbulence suppression at each location. The global effect on $\nu_{\text{eff}}$ is then accounted for through an appropriate phase-space average that depends on the pump mode structure. To qualitatively describe the suppression process, and the resulting feedback on the pump amplitude through  $\nu_{\text{eff}}$, we consider here an electric field averaged over the radial structure based on the ZM amplitude $A_z(t)$ (defined in Eq.~(\ref{amplitudeZ})) as $E_{z,avg}^2 = A_z^2(t)/\int  d \mathcal{V}  \, |\alpha_z(r)|^2$. The constant shear $\gamma_E$ then reads:
\begin{equation}\label{gamma_E}
    \gamma_E = \frac{A_z(t)}{r_s B_0(\int  d \mathcal{V}  \, |\alpha_z(r)|^2)^{1/2}},
\end{equation}
where $r_s$ is the radial location of the pump mode peak and $B_0$ toroidal field.

\begin{figure}[h!]
    \centering
    \includegraphics[width=0.98\linewidth]{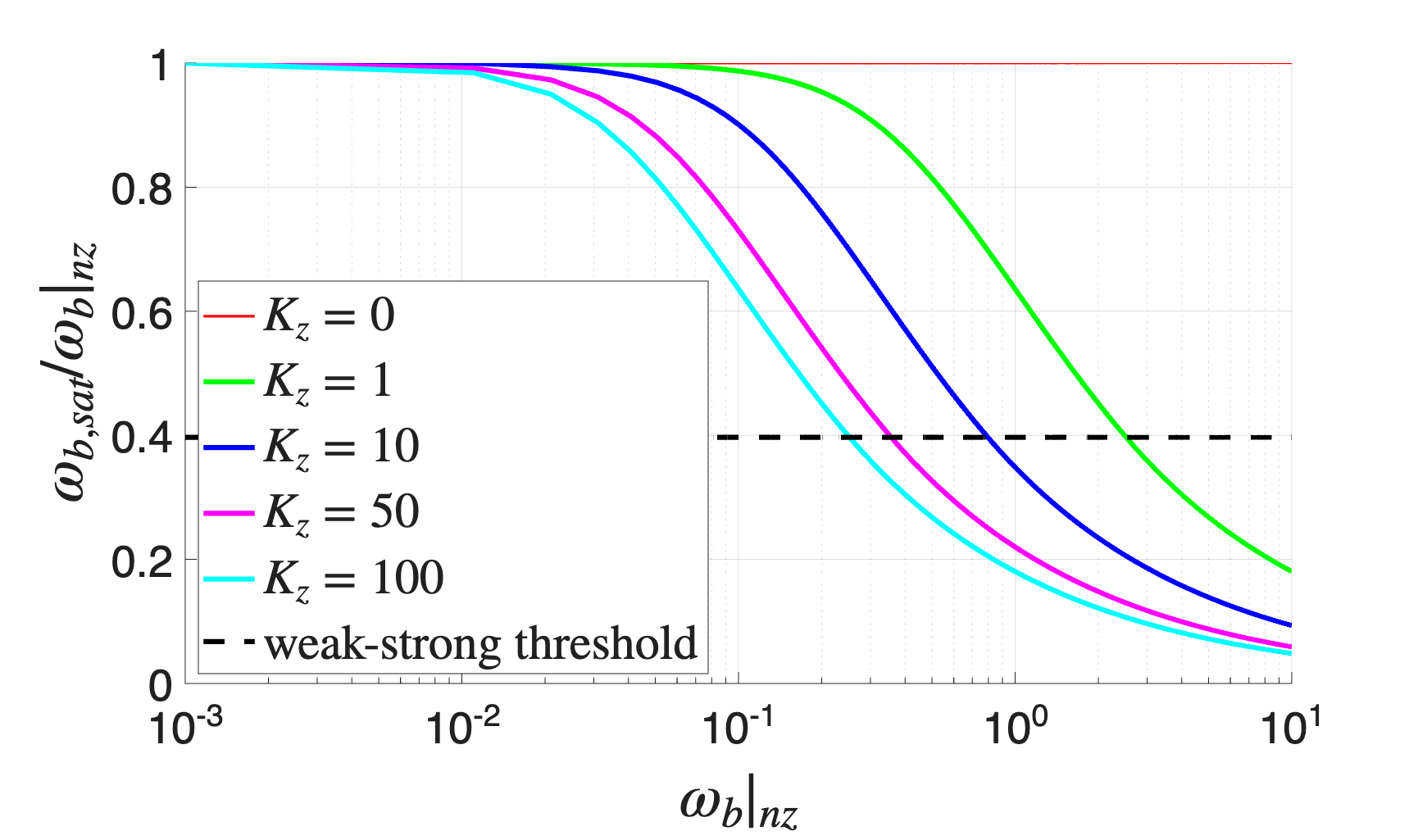}
    \caption{Reduction in the saturated amplitude due to ZM suppression of $\nu_{\text{eff}}$ as a function of the expected saturation in the absence of ZM for different values of $K_z$. The other parameters values are set to $\gamma_c = r_s = B_0 = (\int  d \mathcal{V}  \, |\alpha_z(r)|^2)^{1/2} = 1$. The threshold between weak and strong shearing regimes corresponds to $\omega_{b,sat}/\omega_{b}|_{nz} = 2^{-2/3}$.}
    \label{fig:collisional}
\end{figure}

From the classical works on $\boldsymbol{E}\times\boldsymbol{B}$ stabilization \cite{Biglari_1990, Burrell_1997,Terry_2000}, the shearing flow is expected to suppress the turbulence when the shearing rate $\gamma_E$ is comparable with the linear growth rate of the instability causing the turbulence.
According to the analysis of Ref.\cite{Ivanov_2025}, two regimes for turbulence suppression due to flow shear can be distinguished: "weak" and "strong" shear. Following its notation, a threshold can be qualitatively identified at $\gamma_E\sim \gamma^o$. Here $\gamma^o$ is the linear injection rate, associated with the temperature-gradient instability growth rate in absence of shear, being determined by the balance with turbulence nonlinear mixing \cite{Barnes_2011,Adkins_2022}. 
It is possible to define a critical shear value $\gamma_c \equiv \mathcal{A}^o\gamma^o$, with $\mathcal{A}^o$ being the elongation of turbulent eddies in absence of shear. In our model, we will assume $\mathcal{A}^o\approx 1$ so that $\gamma_E = \gamma_c$ represents the critical shear value for the transition between weak and strong shear suppression regimes described in Ref.\cite{Ivanov_2025}. An accurate description for $\gamma_c$ is beyond the scope of the present work. It will be used as a model parameter that will account also for possible proportionality constants not accounted for in the phenomenological description of Ref.\cite{Ivanov_2025}. 
Under this framework, one can then obtain relations to describe how the thermal ion diffusivity $D_{th,i}(\gamma_E(t))$ can be affected in presence of a shear flow, both in the weak and strong shear regimes:
\begin{align}
    \frac{D_{th,i}(\gamma_E(t))}{D_{th,i}(0)} \sim \left[1+ \frac{A_z(t)}{r_sB_0(\int  d \mathcal{V}  \, |\alpha_z(r)|^2)^{1/2}}\frac{1}{\gamma_c}\right]^{-2} \label{weakshear}
\end{align}
for $\gamma_E<\gamma_c$, and
\begin{align}
    &\frac{D_{th,i}(\gamma_E(t))}{D_{th,i}(0)} \sim \frac{\gamma_c}{4}\left[\frac{A_z(t)}{r_sB_0(\int  d \mathcal{V}  \, |\alpha_z(r)|^2)^{1/2}}\right]^{-1} \label{strongshear}
\end{align}
for $\gamma_E>\gamma_c$. Continuity has been enforced and when $\gamma_E$, defined in Eq.~(\ref{gamma_E}), coincides with $\gamma_c$, then $D_{th,i}(\gamma_c)/D_{th,i}(0) = 1/4$.
Using the fact that $\nu_{turb} \propto D_{EP}^{1/3} \propto D_{th,i}^{1/3}$, we can finally describe how the scattering due to microturbulence changes in time due to the beat-driven ZM:
\begin{equation}\label{nuoft}
    \nu_{\text{eff}}(t) = \nu_{\text{eff}}(t_0)\left[\frac{D_{th,i}(\gamma_E(t))}{D_{th,i}(0)}  \right]^{1/3}
\end{equation}
where $\nu_{\text{eff}}(t_0)$ is the effective scattering due to microturbulence at $t=t_0$, i.e. in absence of ZM induced shearing. 
If the effect on $\nu_{\text{eff}}(t)$ is the only modification in the presence of ZM within the time-local regime, then the nonlinear growth rate can still be described with Eq.~(\ref{nonlinearg}), (as shown in \ref{APPA}) and the saturation amplitude satisfies Eq.~(\ref{nozonal_sat}) modified by $\nu_{\text{eff}}(t)$:
\begin{align}\label{zonal_sat}
    \omega_{b,sat} = \omega_{b}|_{nz} \left[\frac{D_{th,i}(\gamma_E)}{D_{th,i}(0)}  \right]^{1/3}
\end{align}
where we define 
\begin{equation}
    \omega_{b}|_{nz} = \sqrt[4]{\frac{6e^{i\phi }}{\Gamma(1/3)(3/2)^{1/3}}}\left(1-\frac{\gamma_{d,p}}{\gamma_L}\right)^{1/4}\nu_{\text{eff}}(t_0),
\end{equation}
as the saturation expected without ZM effect. Neglecting again the ZM damping rate, $\gamma_{d,z} = 0$, and using Eqs.~(\ref{weakshear}) and (\ref{strongshear}) in Eq.~(\ref{zonal_sat}), the saturation levels are determined by:
\begin{align}
    &\omega_{b,sat}\left[\frac{\omega_{b,sat}^4K_{z}}{\gamma_cr_sB_0(\int  d \mathcal{V}  \, |\alpha_z(r)|^2)^{1/2}} +1\right]^{2/3} =\omega_{b}|_{nz}
    \label{weaksat}
\end{align}
for $\gamma_E<\gamma_c$, and
\begin{align}
    &\omega_{b,sat}=\omega_{b}|_{nz}^{3/7}\left[\frac{\gamma_cr_sB_0(\int  d \mathcal{V}  \, |\alpha_z(r)|^2)^{1/2}}{4K_{z}} \right]^{1/7} 
    \label{strongsat}
\end{align}
for $\gamma_E>\gamma_c$, where we used the beat-driven relation of Eq.~(\ref{forceK}) to write $A_z$ in terms of the pump amplitude.
In Fig.~\ref{fig:collisional} the reduction for the saturated amplitude coming from the solution of Eqs.~(\ref{weaksat}) and (\ref{strongsat}) is plotted against the expected saturation in absence of ZM for different values of $K_{z}$. The threshold between weak and strong shearing regimes $\gamma_E = \gamma_c$ corresponds to a reduction in amplitude $\omega_{b,sat}/\omega_{b}|_{nz} = 2^{-2/3} \approx 0.63$.

While for this analysis we focused on the regime close to threshold, the same mechanism could be used in the regime far from threshold $\omega_b \gg \nu_{\text{eff}}$ leading to similar results \footnote{The nonlinear growth rate far from threshold can be approximately expressed as: $\gamma_{NL}(t)/\gamma_L = 1.756\nu_\text{eff}^3 /\omega_b^3(t)$ \cite{Devin_inprep}}. The ordering far from threshold requires that $\gamma_{d,p} \ll \gamma_L$. In such conditions, characterized by very small background damping compared to the linear growth rate, the effects of finite amplitude ZMs on the sink term should be discussed more carefully since the additional damping from shearing or q profile modifications ~\cite{Spong_2021} may become comparable to $\gamma_{d,p}$.

To conclude this section, we note that we have neglected the ZM collisional damping rate, $\gamma_{d,z}$. This simplification allows us to focus on developing qualitative insights into how self‑generated ZMs impact the saturation amplitude, rather than on quantitative predictions that are beyond the scope of our reduced formalism. Nonetheless, here we briefly discuss potential implications of including $\gamma_{d,z}$, leaving a more detailed treatment of the fully coupled system to future work. If $\gamma_{d,z}$ becomes comparable to the net linear growth rate of the pump wave, $\gamma_L - \gamma_{d,p}$, the ZM will be unable to grow to sufficient amplitude to significantly influence the pump evolution or saturation; in this regime, energy is still transferred from the pump to the ZM, but is rapidly dissipated. Conversely, if $\gamma_{d,z}$ is sufficiently small, the ZM can grow to sufficient levels to affect the pump in the ways discussed here. The impact of $\gamma_{d,z}$ may become more relevant near saturation, when $\gamma_{NL}(t) - \gamma_{d,p} \approx 0$. However, under these conditions also the beat‑driven assumption may break down, and additional nonlinear effects, such as modulational instability, could become significant. 

\section{BOT simulation results}\label{BOT}
In sections \ref{collisionless} and \ref{collisional} we described analytic models to estimate the saturated amplitude of the pump wave in presence of self-generated ZM. In the collisionless regime, due to the finite energy available to the system, we could determine the saturation amplitude without explicit knowledge of $\gamma_{NL}$. 
In the presence of sources and sinks, the same procedure is not possible and in order to proceed analytically we restricted our calculation assuming close to threshold and scattering dominated saturation. Within those assumptions $\gamma_{NL}$ becomes time local and the evolution of the mode amplitude can be treated analytically. In general cases, however, analytical progress cannot be achieved.

In this section, the kinetic problem for the nonlinear growth rate is solved numerically using the BOT code~\cite{Lillley_2010}. It addresses nonlinear wave–particle resonant interactions within the simplified bump-on-tail geometry. As discussed in Refs.~\cite{BerkPPR1997,Berk_2012}, the bump-on-tail paradigm is sufficient to capture the essential physics required to understand the destabilization and saturation of isolated modes via wave–particle nonlinearities for the case of low frequency modes in axisymmetric devices.
The BOT code solves the nonlinear bump-on-tail problem by evolving the bounce (trapping) frequency, that within the code normalization follows $\omega_b/\gamma_L = A_p^{1/2}$, and the particle distribution function self-consistently.
Within BOT, we include the nonlinear generation of ZMs following the coupled equations~(\ref{eq:amplitudeP3}–\ref{eq:amplitudeZ3}). In this implementation, the ZM acts as an additional damping mechanism on the pump wave depending on the evolving amplitude $A_z(t)$. It is important to note that within the code there is no explicit knowledge of radial structures and of $A_z$ being associated to a ZM. Any direct impact that ZMs might have on the particle distribution function is also not included. The only retained effects are the evolution of $A_z$, computed according to our energy conservation arguments, thus reducing the pump growth, and the suppression of the effective scattering rate $\nu_{\text{eff}}$ due to a turbulence reduction associated to $A_z(t)$.

\subsection{Collisionless case}\label{sub_botcolles}

We begin by analyzing the collisionless case, in which the only additional effect considered is the presence of nonlinearly generated wave introduced as an additional damping term in the pump wave evolution. 
When simulating the self-generated wave dynamics, the initial condition is not arbitrary but must satisfy the beat-driven relation given in Eq.~(\ref{forceK}). The parameter $K_{z}$, which links the amplitude of the self-generated ZM to the squared amplitude of the pump wave, varies depending on the characteristics of the mode and the specific plasma parameters. It has the dimensions of the inverse of $A_p$, and is thus normalized accordingly.
However, our aim here is not to provide a quantitatively accurate prediction but rather to illustrate the qualitative impact of ZM generation. Thus, we treat $K_{z}$ as an input parameter that effectively controls the strength of beat-driven ZM generation.  We compare the numerical results with analytic saturation estimates from Eq.~(\ref{quintic2}) using $K_{hh} = K_{z}$. It is worth noting that, while for the analytic case of the 1D model we had to consider the toy model involving the beat-driven higher harmonic, the code solves the coupled Eqs.~(\ref{eq:amplitudeP3}) and (\ref{eq:amplitudeZ3}) directly.

\begin{figure}[h!]
\centering
\begin{subfigure}{\linewidth}
  \centering
  \includegraphics[width=0.98\linewidth]{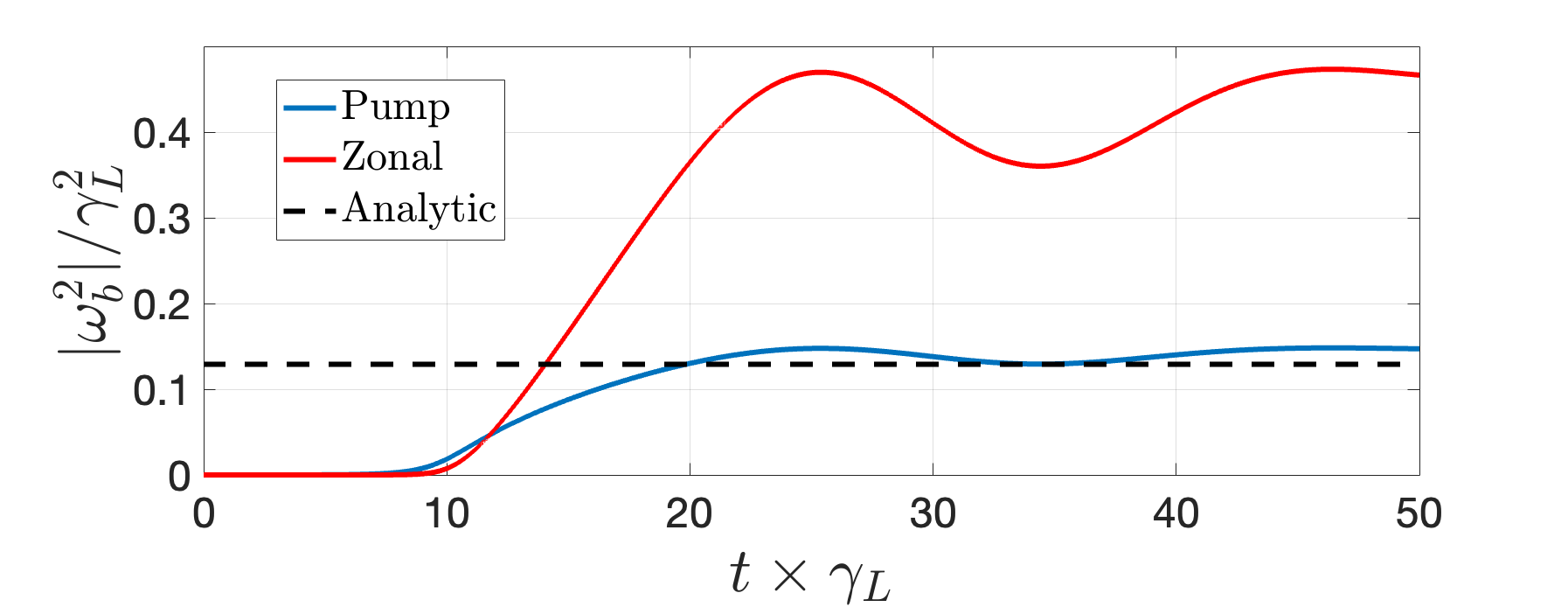}
  \caption{}
  \label{K22}
\end{subfigure}
\\
%\hfill
\begin{subfigure}{\linewidth}
  \centering
  \includegraphics[width=0.98\linewidth]{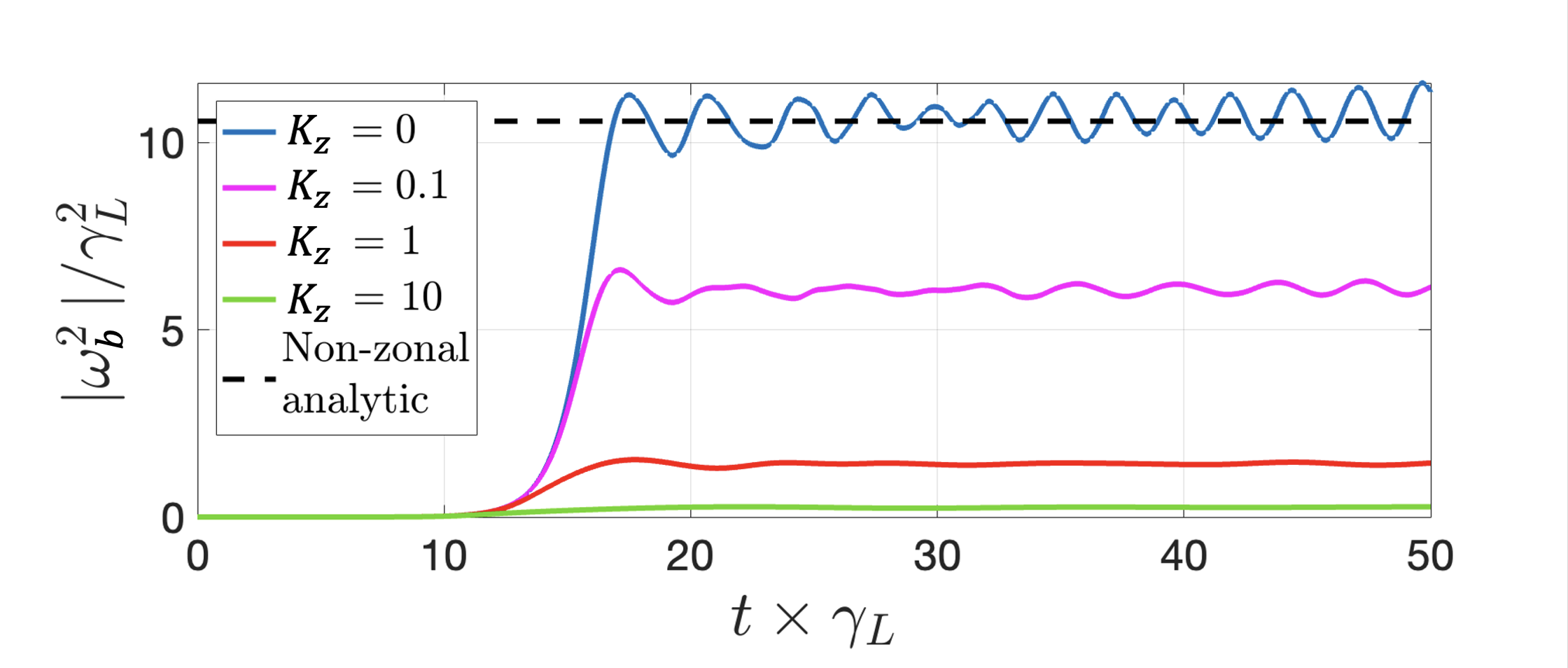}
  \caption{}
  \label{colless_Kscan}
\end{subfigure}
\caption{In part (a), the normalized amplitude evolution of pump and associated beat-driven ZM $A_z(t) = K_{z}A_p^2(t)$ are plotted for $K_{z} =22$. The analytic solution of Eq.~(\ref{quintic2}) for the expected saturation is showed as dashed black line. Part~(b) shows the evolution of pump wave amplitudes for different values of the parameter $K_{z}$. The corresponding ZM amplitude, following the quadratic relation for each $K_{z}$ value, is not shown. The expected saturation for the non-zonal case with $\omega_b/\gamma_L \approx 3.2$, is showed as dashed line. The initial condition for the pump wave amplitude is $\omega_b^2/\gamma_L^2=10^{-6}$.}
\label{collisionless_BOT}
\end{figure}

The results for a range of $K_{z}$ values are reported in  Fig.~\ref{collisionless_BOT}. In Fig.~\ref{K22}, BOT results are compared with the analytic prediction for the saturation coming from Eq.~(\ref{quintic2}) using  $K_{z}=22$, corresponding to the case showing agreement between the quintic equation solution and gyrokinetic simulations in Fig.~\ref{fig:comparison}. The analytic model slightly underpredicts the normalized saturation amplitude. It is important to remark that for the entirety of the simulation the beat-driven relation of Eq.~(\ref{forcedriven}) is assumed. Thus, the ZM is following the quadratic relation with the pump even after the first amplitude peak. This is a consequence of our assumptions and, if the ZM amplitude becomes large enough, other nonlinear mechanisms which are not accounted for here might become dominant. Nevertheless, the model can, at least partially, capture the reduction in saturation amplitude due to self-generation of ZM, and, as shown in Fig.~\ref{colless_Kscan}, the reduction is stronger when the parameter $K_{z}$ is increased. The case with $K_{z} = 0$ is reported in Fig.~\ref{colless_Kscan} together with the expected non-zonal saturation value of $\omega_b/\gamma_L \approx 3.2$ \cite{Fried_1971}.

\subsection{Collisional case}\label{sub_botcoll}

\begin{figure}
  \centering
  \includegraphics[width=0.95\linewidth]{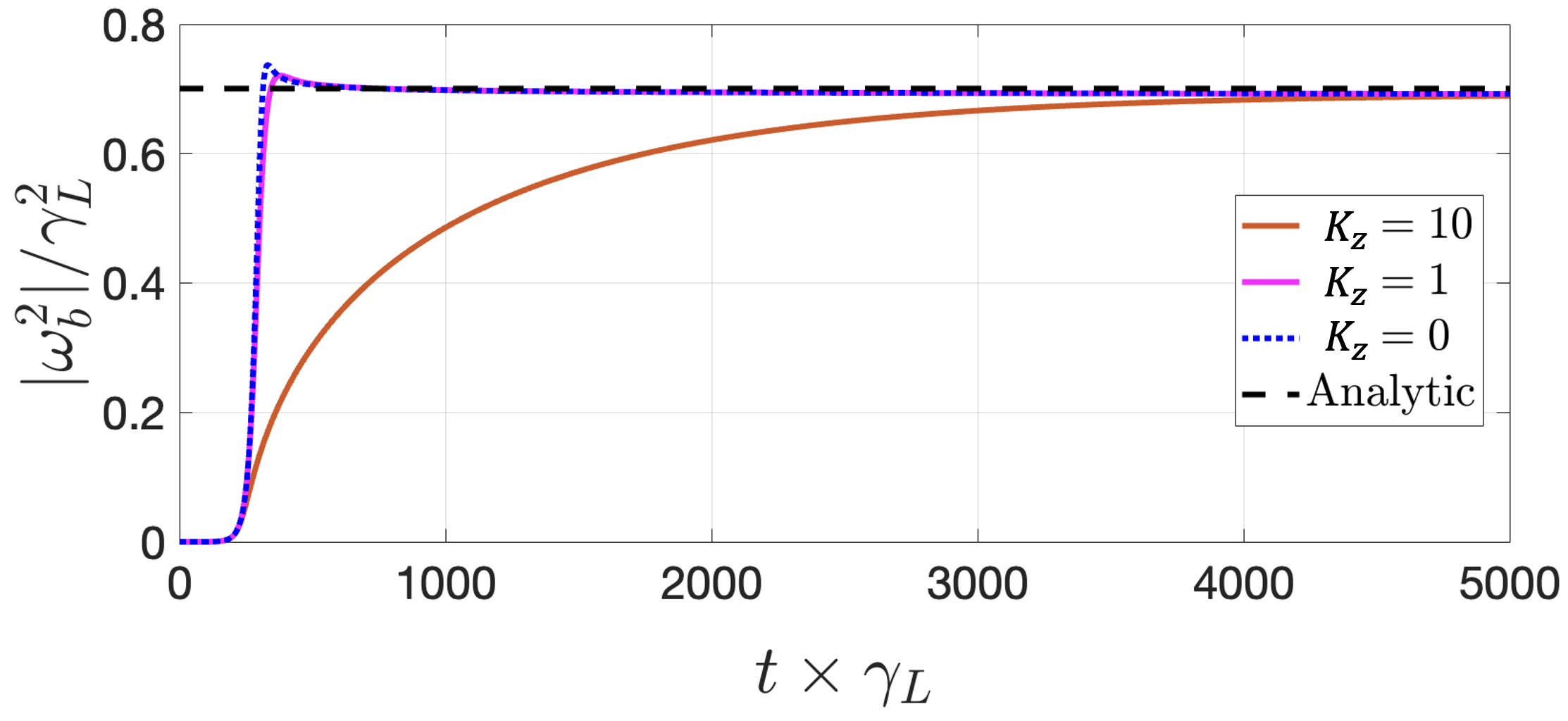}
  \caption{Evolution of pump wave amplitude close to threshold for different values of $K_{z}$ compared with the saturation predicted by analytic theory in absence of ZM generation. Sources and sinks parameters used are: $\gamma_{d,p}/\gamma_L = 0.95$ and $\nu_{\text{eff}}/(\gamma_L-\gamma_{d,p}) = 30$. The initial condition for the pump wave amplitude is $\omega_b^2/\gamma_L^2=10^{-6}$.}
  \label{no_nu_marginal}
\end{figure}

As detailed in the previous section, the collisionless case provides valuable insights into how nonlinear wave–wave interactions can reduce mode growth and ultimately lower the saturation amplitude when the energy available to the mode is finite. However, the inclusion of sources and sinks fundamentally changes the system dynamics.
We now present BOT simulation results in the presence of sources and sinks, where the only ZM effect retained is the additional damping introduced by wave–wave generation, implemented under the beat-driven assumption. We first consider a scenario near marginal stability, for which analytic theory predicts monotonic growth followed by steady saturation of the pump wave. In this regime, the expected saturation amplitude in the absence of ZM generation is given by
$ \omega_{b}\bigl|_{nz} \approx 1.18 \bigl(1 -\gamma_{d,p}/\gamma_L\bigr)^{1/4} \, \nu_{\mathrm{eff}}$, as derived in Ref.~\cite{BerkPPR1997}. Furthermore, based on the analysis of Sec.~\ref{collisional}, we anticipate that if only ZM generation is included, the final saturation level should remain insensitive to variations in the parameter $K_{z}$. This behaviour is confirmed in Fig.~\ref{no_nu_marginal}.
The evolution of simulations with varying $K_{z}$ shows that ZM generation slows the initial mode growth, delaying the time at which saturation is reached. However, the final saturation amplitude remains unchanged and agrees with the analytic prediction for the case without ZM generation.

\begin{figure}[h!]
\centering
\begin{subfigure}{\linewidth}
  \centering
  \includegraphics[width=0.98\linewidth]{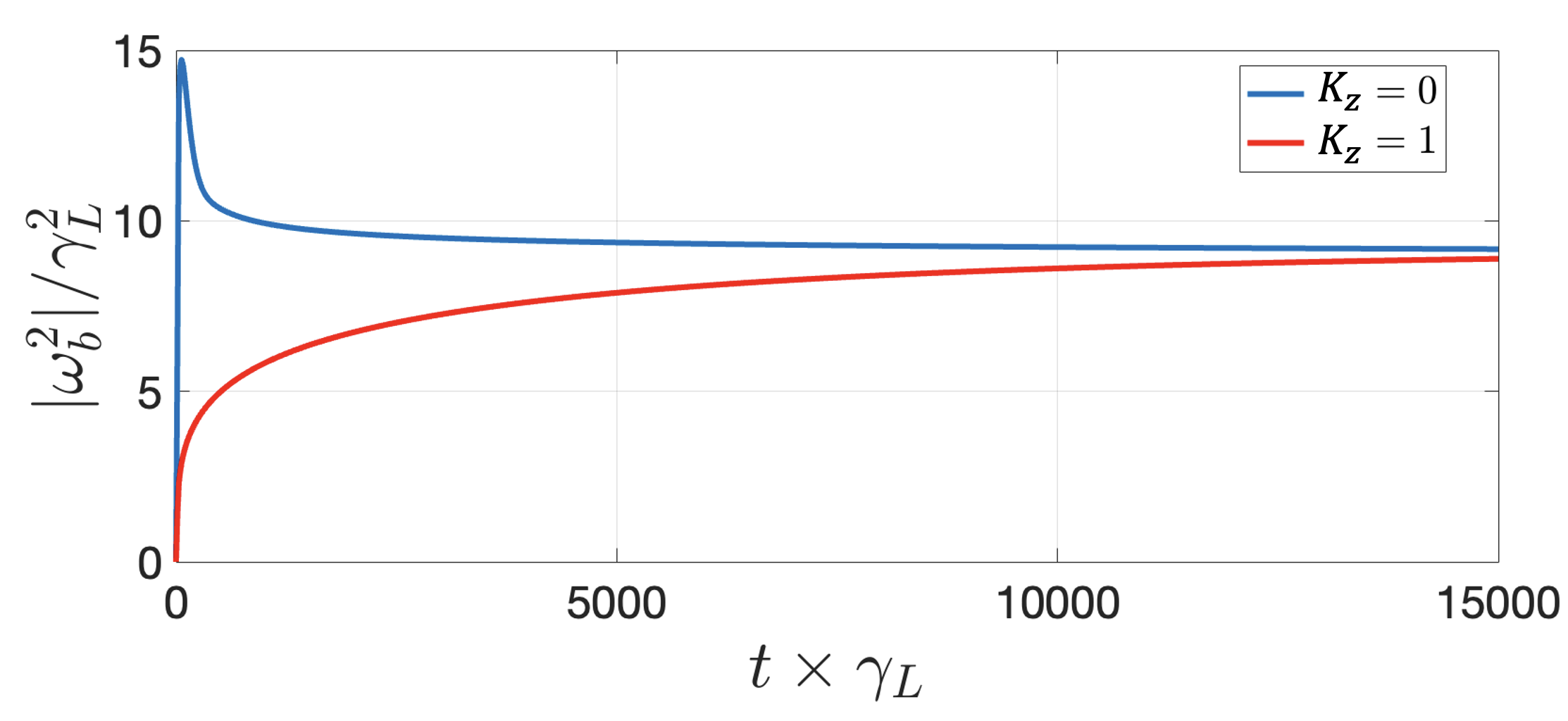}
  \caption{}
  \label{}
\end{subfigure}
\\
%\hfill
\begin{subfigure}{\linewidth}
  \centering
  \includegraphics[width=0.98\linewidth]{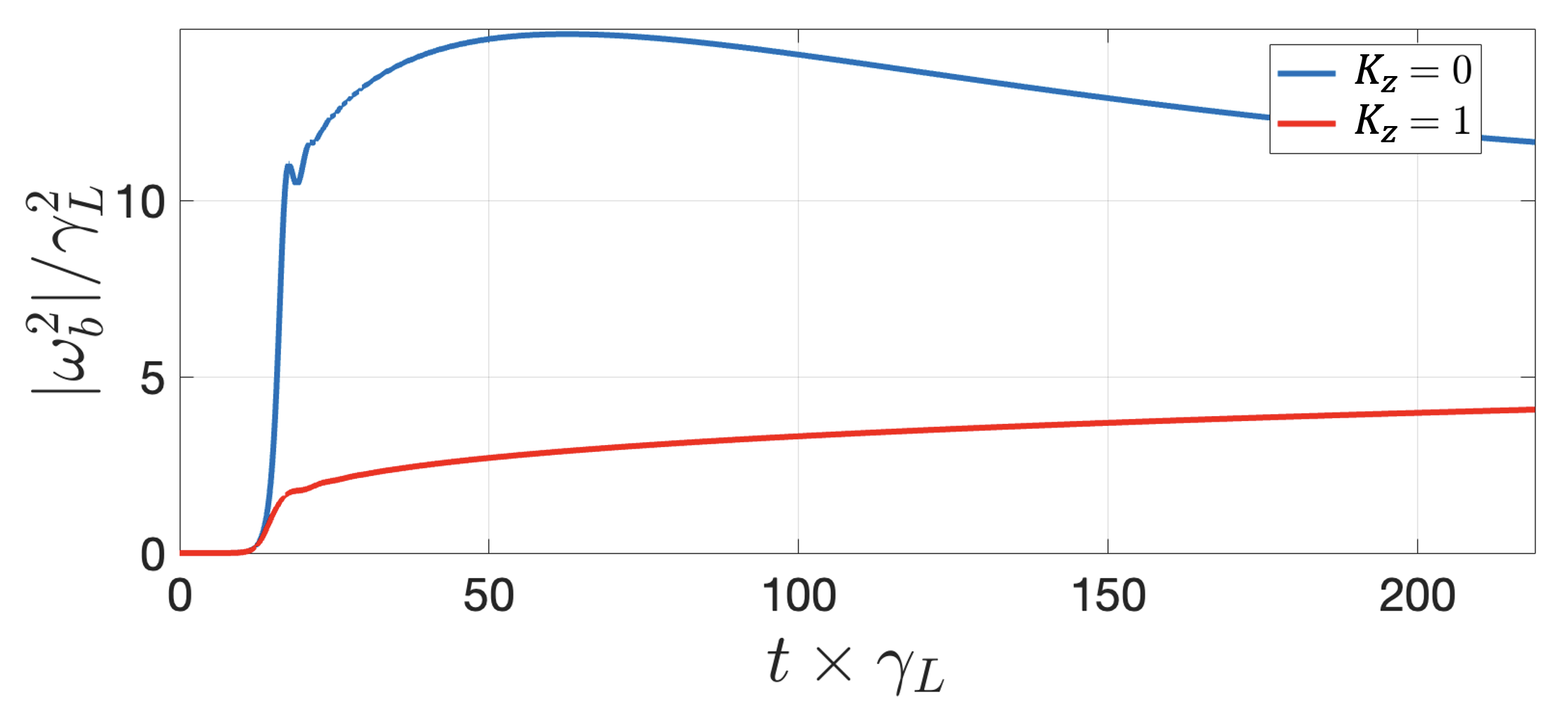}
  \caption{}
  \label{zoom_far}
\end{subfigure}
\\
%\hfill
\caption{Evolution of pump wave amplitude far from threshold with (red curve) and without (blue curve) ZM generation shown in part (a). Zoom around the early exponential growth is shown in part~(b). Sources and sinks parameters used are: $\gamma_{d,p}/\gamma_L = 0.01$ and $\nu_{\text{eff}}/(\gamma_L-\gamma_{d,p}) = 0.55$. The initial condition for the pump wave amplitude is $\omega_b^2/\gamma_L^2=10^{-6}$.}
\label{no_nu_far}
\end{figure}

This behaviour is not limited to marginal stability case alone. As shown in Fig.~\ref{no_nu_far}, the same insensitivity to beat-driven generation is observed far from marginal stability, for a region of parameters for which theory predicts a steady saturation \cite{BB1990_1,BB1990_2,Berk_1990_3}. In these cases, ZM generation strongly alters the early “collisionless” growth phase by reducing the energy available from the distribution function. However, when considering saturation in the presence of sources and sinks, one must examine amplitude evolution on collisional and dissipative time scales. As collisions act to restore the distribution function and balance background damping, the final saturated amplitude converges to the same value regardless of ZM generation. The BOT simulations corroborate the simplified model’s prediction: if beat-driven ZM generation is the sole additional effect, then the saturation level of the pump wave set by sources and sinks remains unaffected by ZM dynamics.

\begin{figure}[h!]
\centering
\begin{subfigure}{\linewidth}
  \centering
  \includegraphics[width=0.98\linewidth]{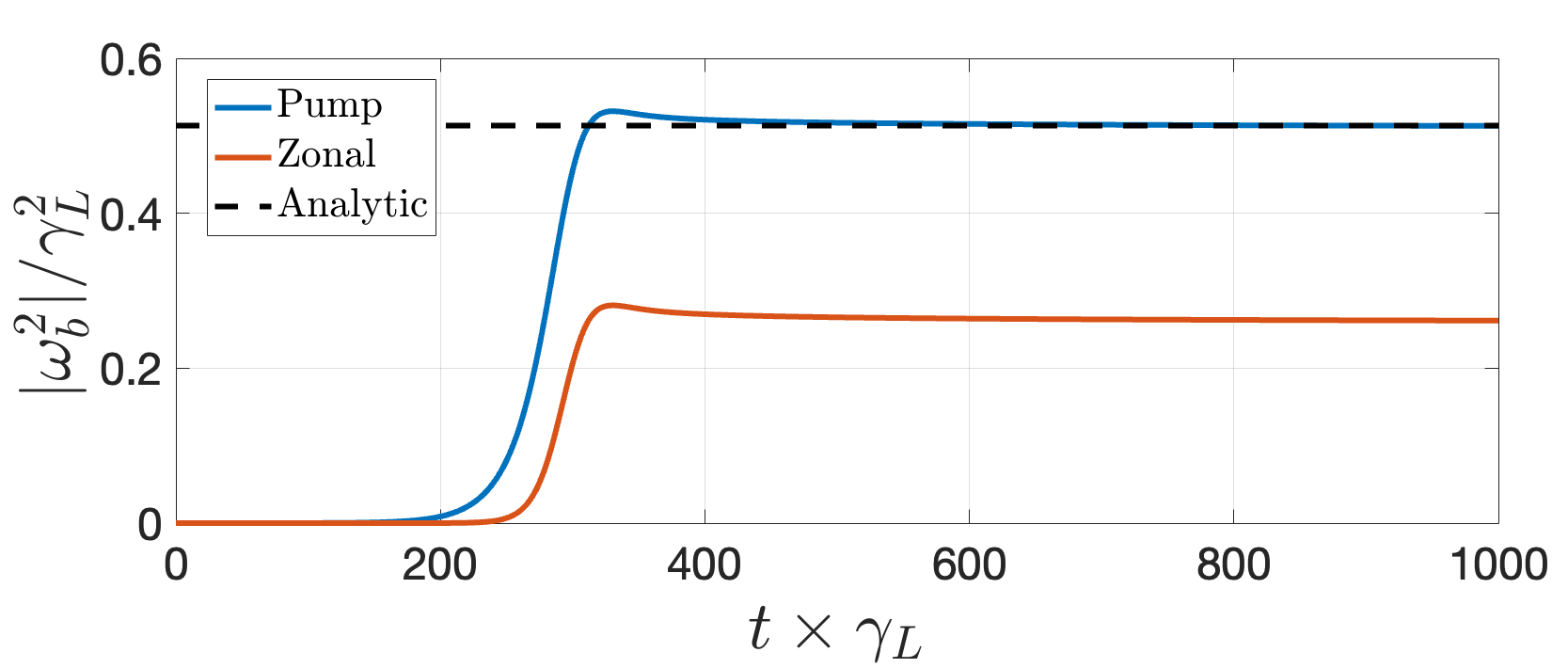}
  \caption{}
  \label{coll_nu_K_1}
\end{subfigure}
\\
%\hfill
\begin{subfigure}{\linewidth}
  \centering
  \includegraphics[width=0.98\linewidth]{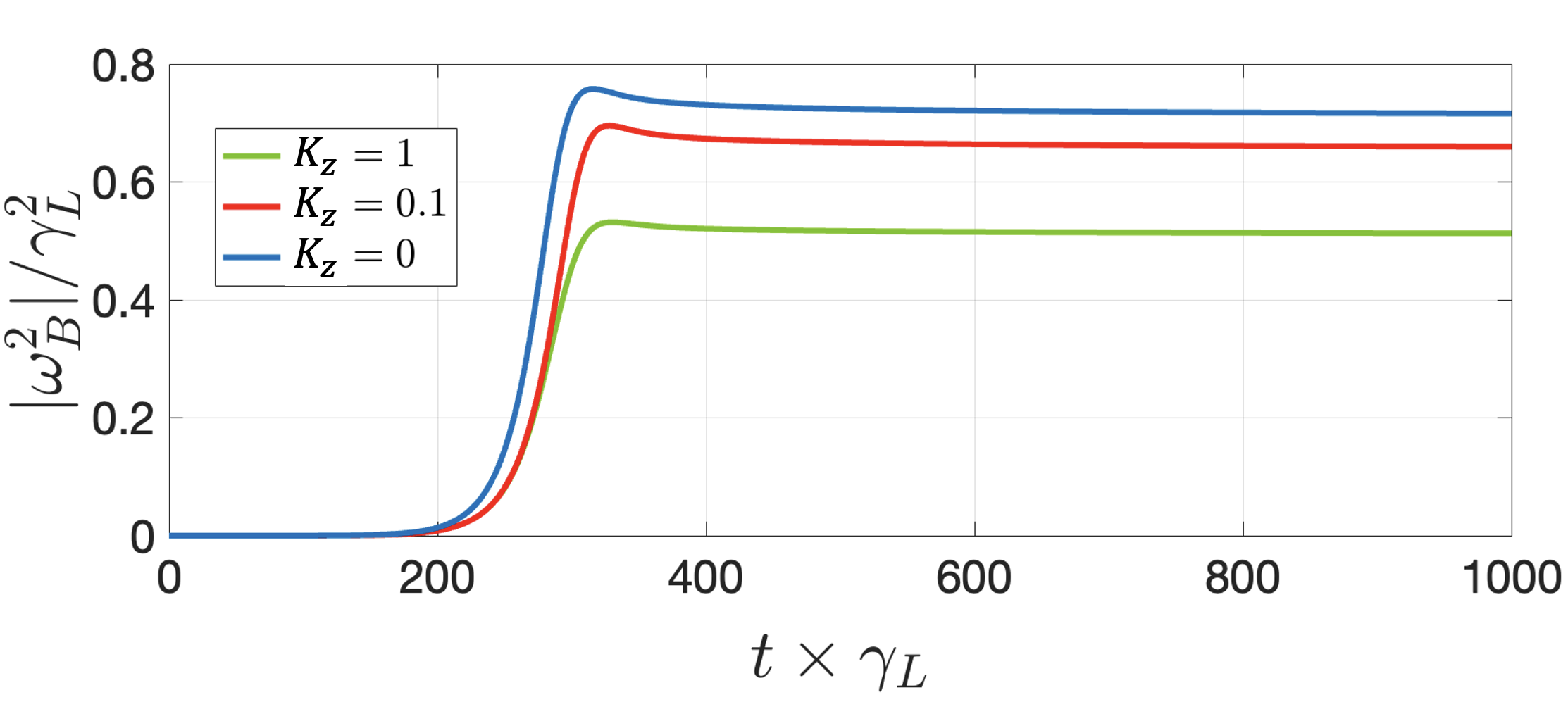}
  \caption{}
  \label{coll_nu_K_scan}
\end{subfigure}
\caption{In part (a) the normalized amplitude evolution of pump and associated beat-driven ZM are plotted for $K_{z} =1$. The analytic solution of Eq.~(\ref{weaksat}) for the expected saturation shown as a dashed black line. Part (b) shows the evolution of pump wave amplitudes for different values of the parameter $K_{z}$. The corresponding ZM amplitude, following the quadratic relation for each $K_{z}$ value, is not shown. Sources and sinks parameters used are the same ones of Fig.\ref{no_nu_marginal}: $\gamma_{d,p}/\gamma_L = 0.95$ and $\nu_{\text{eff}}/(\gamma_L-\gamma_{d,p}) = 30$. The initial condition for the pump wave amplitude is $\omega_b^2/\gamma_L^2=10^{-6}$.}
\label{coll_nu_t}
\end{figure}

Following the analytic discussion, one mechanism by which self-generated ZMs can reduce the pump saturation amplitude is through their impact on the source term, i.e., the effective scattering rate $\nu_{\mathrm{eff}}$. In our BOT simulations, we implement a time‐dependent scattering rate that varies with the ZM amplitude, as given by Eq.~(\ref{nuoft}). We focus again on the near‐marginal‐stability regime, where theory predicts a low‐amplitude saturation. For small values of the parameter $K_{z}$, the turbulence reduction due to the ZM effect is governed by the “weak shear” suppression formula in Eq.~(\ref{weaksat}), as shown also in Fig.~\ref{fig:collisional}. We first present, in Fig.~\ref{coll_nu_K_1}, the evolution obtained with $\nu_{\mathrm{eff}}(t)$ for a representative value of $K_{z} = 1$ together with the corresponding beat‐driven ZM amplitude. We then perform a parameter scan in Fig.~\ref{coll_nu_K_scan}, which shows that increasing $K_{z}$ again leads to a stronger reduction of the saturated pump amplitude.

\section{Conclusion}\label{Conclusions}

This manuscript presents a simplified and transparent framework for evaluating how, at least qualitatively, wave–wave nonlinearities can modify the evolution and saturation of Alfv\'enic modes driven by wave–particle resonances. From a practical standpoint, it is well known that the saturation amplitude of EP-driven instabilities is crucial to understand their associated EP transport and losses. 
Wave–particle nonlinear models have been highly successful in predicting the saturation of EP-driven instabilities under a wide range of conditions \cite{Breizman_2011}. However, as discussed in the introduction, experimental evidence, for example the TFTR cases described in Ref.~\cite{Todo_2010}, shows that for sufficiently large-amplitude modes, the saturation predicted by these models can be overestimated, as wave–wave nonlinear effects become significant. This implies that experimental modeling and predictive simulations must, in such cases, rely on more complete and computationally costly tools. The simplified model presented here is intended to mitigate this limitation by extending the predictive capabilities of significantly cheaper wave-particle-based frameworks. Such an approach could be particularly valuable during scenario development for future devices, where broad parameter scans for optimization may be prohibitively expensive with high-fidelity simulations. In addition, the model yields predictions for the amplitude of beat-driven ZMs, which could be tested against experimental measurements when available, providing further opportunities to benchmark and stress-test the reduced model capabilities. 

Extending the standard amplitude‐evolution equation for wave–particle interactions, we introduced an additional damping term to account for nonlinear wave generation. We specifically examined the beat‐driven excitation of zero‐frequency $(m,n)=(0,0)$ zonal modes, a mechanism observed in multiple gyrokinetic simulations and supported by analytic theory, where the ZM grows at approximately twice the rate of the primary “pump” wave.

Although this beat‐driven process is only one of the possible wave–wave nonlinear effects, we argue that it represents the leading‐order effect during the linear and early nonlinear growth stages. Focusing on it allows us to isolate the primary influence of ZMs on the pump wave saturation amplitude. Rather than attempting to reproduce the full complexity of wave–wave nonlinearities, our simplified model tries to clarify the key role of ZM generation in energy transfer and saturation dynamics. 

Our analyses highlight that, in the collisionless limit, the generation of beat-driven waves alone significantly alters both the growth and saturation of the pump mode. Simple energy-conservation arguments reveal that no additional ZM specific physics is required to predict a reduction in the pump saturation amplitude: energy that would have been transferred from the fast-ion distribution to the pump is instead redirected into the beat-driven modes. Consequently, not only is the available energy distributed between the pump and the beat-driven modes, but the pump capacity to extract further energy from the distribution is diminished. By introducing a single free parameter to quantify the beat-driven generation, our reduced model reproduces the trends in saturation amplitude observed in more comprehensive gyrokinetic simulations in this collisionless regime.

The more realistic scenario, including sources and sinks, behaves quite differently. In the presence of particle scattering and background damping, the energy available to drive the pump mode is constantly replenished, so beat-driven wave generation can slow down growth but does not by itself affect the saturation level. Under these conditions, the specific properties of the ZM become critical in determining its influence on the pump evolution. Typically, the collisional damping rate of the ZM is small compared to its beat‐driven growth rate \cite{Diamond_2005}, at least during the pump linear growth phase, allowing the ZM amplitude to grow sufficiently to impact the pump wave through various channels. 
In this work, we have neglected potential modifications to the pump background damping or to the wave–particle resonance itself, and instead focused on the ZM’s effect on particle scattering via turbulence suppression. By reducing microturbulence, the ZM effectively diminishes the resonant‐particle source that replenishes the distribution through scattering of nonresonant particles into the resonant island. This mechanism lowers the saturation amplitude on the collisional timescale, in addition to the growth‐rate reduction caused by energy transfer into the ZM.  

This result underscores the necessity of properly accounting for sources and sinks when studying the evolution and saturation of modes driven by wave–particle resonances. As noted in earlier works \cite{zakharov1962,Su_Oberman,BB1990_1,Berk_1996,Callen2014PoPCoulombCollsions,Duarte_2019,Catto_2021,Hamilton_2023}, within resonances the effect of scattering is exacerbated, extending the resonance interaction range. Here, we emphasize that wave–wave nonlinearities also affect the system in a fundamentally different way in the presence of continuous particle sources and sinks. While the collisionless case provides valuable insight into how initial, exponential growth is modified by wave–wave interactions, the lack of sources and sinks leads to a qualitatively different saturation regime.

The model presented here is highly simplified and does not capture many important effects that may dominate the influence of wave–wave nonlinearities on the evolution and saturation of Alfv\'enic modes driven by wave–particle resonances, such as modulational instability \cite{Chen_Zonca_2012,Qiu_2023} and other feedback mechanisms that the ZM can have on the pump wave \cite{Spong_1994,Brochard_2024,Chen_2025}. On the other hand, by developing our model within the standard framework used for nonlinear wave–particle dynamics, and thanks to its simple formalism, it can be readily implemented in existing codes based on perturbative models. While still far from a complete treatment of the fully nonlinear problem, augmenting a reduced wave–particle code with the mechanism proposed here allows, at least partially, for inclusion of nonlinear wave–wave effects, specifically the beat‑driven generation of ZMs, on mode evolution and saturation.

We also point out that an informative reduced approach for the coupling between fast-ion-driven modes, zonal modes and turbulence has been recently proposed by Yan and Diamond in \cite{Yan_2025_arXiv}. Their model focuses on the characteristics of a beat-driven ZM and how it can couple directly to the turbulence through a process known as geodesic acoustic transfer. The pump mode evolution is, however, limited to a constant linear growth in the absence of ZMs, not capturing details of the wave-particle nonlinearities. The reduced model presented here, instead, focuses on the pump wave evolution and consequent fast ion transport, highlighting the role of the ZM in affecting the saturation that can be achieved due to wave-particle nonlinearities. Although the two proposed models try to address similar aspects of the fast-ion-driven modes, zonal modes and turbulence interactions, they focus on different key features of the problem and can be related in a complementary way.

\section*{Acknowledgments}

We acknowledge several stimulating discussions during the 18th Technical Meeting on Energetic Particles in Magnetic Confinement Systems in Seville in March 2025, where this work was first presented. In particular, we thank  M. V. Falessi, M. Fitzgerald, B. N. Breizman, Z. Qiu, and J. Ruiz Ruiz.  TB thanks W. W. Heidbrink, X. D. Du, and the attendees of the DIII-D early career fast ion meetings, where this work was discussed.

We thank T. Adkins for drawing our attention to recent advances in quantifying turbulence suppression via shearing \cite{Ivanov_2025}, J. P. Lee for his assistance in extending the BOT code, M. K. Lilley for developing and making the BOT code openly available, G. Brochard for sharing and discussing his GTC simulation results, Y. Chen for sharing his GEM simulation results, and F. Zonca, I. Y. Dodin and A. Bierwage for stimulating comments and discussions.

This manuscript is based upon work supported by the US Department of Energy, Office of Science, Office of Fusion Energy Sciences, and has been authored by Princeton University under Contract DE-AC02-09CH11466 with the US Department of Energy (DOE). The work was supported by the DOE Early Career Research Program, project \textit{Phase-Space Engineering of Supra-Thermal Particle Distribution for Optimizing Burning Plasma Scenarios}. The publisher, by accepting the article for publication, acknowledges that the United States Government retains a non-exclusive, paid-up, irrevocable, world-wide license to publish or reproduce the published form of this manuscript, or allow others to do so, for United States Government purposes.

\appendix

\section{Validity of the time-local approximation for a time-dependent $\nu_{\text{eff}}$.}\label{APPA}

The nonlinear evolution of EP-driven instabilities in presence of scattering and damping rates can be described analytically by the Berk-Breizman cubic equation, introduced in Ref.\cite{Berk_1996}. This time-delayed integro-differential equation contains rich nonlinear features, and has been applied to interpret a variety of experimental phenomena including oscillations near the saturation level, pitchfork splitting, chaotic mode evolution, and the emergence of frequency chirping. 
As shown in Refs.\cite{Duarte_2019NF,Lestz_2021}, further simplification is possible in scattering dominated regimes, in which the cubic equation becomes time-local. Previous studies, however, considered a constant $\nu_{\text{eff}}$. In this Appendix, we derive the condition for which the time-local approximation holds in the presence of a time dependent $\nu_{\text{eff}}(t)$. We start from the 1D kinetic equation for the distribution function $f$ in normalized spatial and velocity coordinates, $\xi$ and $\Omega$, 

\begin{equation} \label{app:kinetic}
\frac{\partial f}{\partial t} + \Omega \frac{\partial f}{\partial \xi} + Re\left[\omega_b^2e^{i\xi}\right]\frac{\partial f}{\partial \Omega} = \nu^3_{\text{eff}}(t)\frac{\partial^2 (f-F_0)}{\partial \Omega^2}
\end{equation}
and the power balance equation,

\begin{align} \label{app:powerB}
\nonumber & P = -\frac{e\omega}{k}\int d\xi d\Omega E(\xi,t)f(\xi,\Omega,t)  \\ & = \frac{d}{dt}\int d\xi\frac{E^2}{4\pi} + 2\gamma_{d,p} \int d\xi\frac{E^2}{4\pi}
\end{align}
In the kinetic equation (\ref{app:kinetic}), $\omega_b$ is the bounce frequency, $\nu_{\text{eff}}(t)$ is the effective scattering frequency, and $F_0$ is the equilibrium distribution in the absence of the mode. In the power balance Eq.~(\ref{app:powerB}), $e$ is the elementary charge, $\omega$ the resonant frequency, $k$ the wavenumber, $E$ the electric field amplitude, and $\gamma_{d,p}$ the background damping rate. Following \cite{Berk_1996} and \cite{BerkPPR1997}, we then express $f$ as a Fourier series in $\xi$,

\begin{equation} \label{fourierexp}
f(\xi,\Omega,t) = F_0(\Omega) + f_0(\Omega,t) + \sum_{l = 1}^{\infty}\left(e^{il\xi}f_l(\Omega, t)\right) + c.c.,
\end{equation}
and expand perturbatively in the small parameter 

\begin{equation} \label{epsparam}
\epsilon \equiv \frac{\omega_b^2}{\nu^2_{\text{eff}}(t)} \ll 1.
\end{equation}
We then take the equation order by order in $\epsilon$ (the order in $\epsilon$ of each term is denoted by the upper index and the Fourier harmonic is denoted by the lower index). The first three orders that contribute to the resulting evolution equation are given by 

\begin{equation} \label{app:f11}
\frac{\partial f_1^{(1)}}{\partial t} - i\Omega f_1^{(1)} = -\frac{1}{2} \omega_b^2(t)\frac{\partial F_0}{\partial \Omega} + \nu_{\text{eff}}^3(t) \frac{\partial^2 f_1^{(1)}}{\partial \Omega^2},
\end{equation}

\begin{equation} \label{app:f02}
\frac{\partial f_0^{(2)}}{\partial t} - \nu^3_{\text{eff}}(t) \frac{\partial^2 f_0^{(2)}}{\partial \Omega^2} = -\frac{1}{2}\left( \omega_b^2(t) \frac{\partial f_1^{(1)*}}{\partial \Omega} + \omega_b^{2*}(t) \frac{\partial f_1^{(1)}}{\partial \Omega} \right),
\end{equation}

\begin{equation} \label{app:f13}
\frac{\partial f_1^{(3)}}{\partial t} - i\Omega f_1^{(3)} = -\frac{1}{2} \omega_b^2(t)\frac{\partial f_0^{(2)}}{\partial \Omega} + \nu_{\text{eff}}^3(t) \frac{\partial^2 f_1^{(3)}}{\partial \Omega^2}.
\end{equation}
It is possible to successively solve Eqs.~(\ref{app:f11})-(\ref{app:f02})-(\ref{app:f13}) by Fourier transforming in $\Omega$ and introducing an integrating factor. Once $f_1^{(1)}, f_1^{(3)}$ are determined, we can substitute them into the power balance equation to solve for the mode amplitude as a function of time. When this is done, the full version of the cubic equation can be obtained \cite{Berk_1996,BerkPPR1997}:

\begin{multline}
\frac{d\omega_b^2}{dt} = (\gamma_L - \gamma_{d,p})\omega_b^2 \\ -\frac{\gamma_L}{2}\int_{0}^{t} d\tau\int_{0}^{\tau}d\tau_1  \left[\omega_b^{2}(\tau)\omega_b^{2*}(\tau_1)\omega_b^{2}(\tau_1+\tau-t)\right](\tau-t)^2 \\ \times
\exp\Bigg(-\int_{\tau_1+\tau-t}^{\tau_1} \nu^3_{\text{eff}}(\tau_3)\left(\tau_1+\tau-t-\tau_3 \right)^2 d\tau_3  \\ -\int_{\tau_1}^{\tau} \nu^3_{\text{eff}}(\tau_3)(t-\tau)^2 d\tau_3 -\int_{\tau}^{t} \nu^3_{\text{eff}}(\tau_3)(\tau-\tau_3)^2 d\tau_3\Bigg) 
\end{multline}
To isolate the leading order effect, we expand $\nu^3_{\text{eff}}(\tau_3)$ as

\begin{equation}
\nu^3_{\text{eff}}(\tau_3) \approx \nu^3_{\text{eff}}(t) + 3\nu^2_{\text{eff}}(t)\frac{d\nu_{\text{eff}}(t)}{dt}(\tau_3-t)
\end{equation}
We may neglect the first-order correction to the integrands exactly when $\nu_{\text{eff}}(t)$ satisfies the adiabaticity condition

\begin{equation}\label{app:adiab}
\frac{1}{\nu_{\text{eff}}(t)}\frac{d\nu_{\text{eff}}(t)}{dt}(\tau_3 - t) \sim \frac{1}{\nu^2_{\text{eff}}(t)}\frac{d\nu_{\text{eff}}(t)}{dt} \ll 1.
\end{equation}

When this adiabaticity condition is satisfied, the time history of $\nu_{\text{eff}}(t)$ can be neglected, then the modified form of the time-delayed cubic equation can be obtained by explicitly integrating the exponential terms:

\begin{align} 
\nonumber &\frac{d\omega_b^2}{dt} = (\gamma_L - \gamma_{d,p})\omega_b^2 - \\ \nonumber &\frac{\gamma_L}{2} \int_{0}^{t/2} d\tau \tau^2 \int_{0}^{t-2\tau}d\tau_1
\omega_b^2(t - \tau)\omega_b^2(t - \tau - \tau_1) \times \\ & \omega_b^{2*} (t-2\tau -\tau_1) 
\exp\left( -\nu^3_{\text{eff}}(t)\tau^2\left(\frac{2}{3}\tau + \tau_1 \right)\right).\label{cubic_time_nu}
\end{align}

Equation (\ref{cubic_time_nu}) is the time-delayed cubic equation obtained by Ref.\cite{Berk_1996}, with the additional time dependence of $\nu_{\text{eff}}(t)$. Further simplification occurs in the time-local regime \cite{Duarte_2019NF}, when

\begin{equation}\label{app:timelocal}
\frac{\gamma_L - \gamma_{d,p}}{\nu_{\text{eff}}(t)} \ll 1.
\end{equation}
In the time-delayed cubic equation, a large effective scattering relative to the mode net growth timescale will cause the integrand to damp very quickly away from $\tau = 0$ and $\tau_1 = 0$. In this regime, the time shifts are unimportant and Eq. (\ref{cubic_time_nu}) is reduced to a Landau-Stuart form:

\begin{equation}
\frac{d\omega_b^2}{dt} = (\gamma_L - \gamma_{d,p})\omega_b^2 - \frac{\gamma_L}{6} \left( \frac{3}{2} \right)^{1/3} \Gamma \left(\frac{1}{3}\right) \frac{\omega_b^2 |\omega_b^2|^2}{\nu_{\text{eff}}^4(t)}
\label{LandauStuart}
\end{equation}

From Eq.~(\ref{LandauStuart}), one can obtain the nonlinear growth rate, Eq.~(\ref{nonlinearg}), from the definition $d\omega_b^2/dt = [\gamma_{NL}(t)-\gamma_{d,p}]\omega_b^2$, as well as the saturation amplitude of Eq.~(\ref{nozonal_sat}) with the only addition of the local time dependence of $\nu_{\text{eff}}(t)$. 

In the remainder of the Appendix, we show that for time-dependent scattering $\nu_{\text{eff}}(t)$, arising from microturbulence suppression due to the beat-driven ZM introduced in Eq.~(\ref{nuoft}), the adiabaticity condition given in Eq.~(\ref{app:adiab}) is satisfied. From Section~\ref{sub:subZFsupp}, recall that the turbulent effective scattering is given by:

\begin{equation} \label{nu_eff_weak_shear}
\nu_{\text{eff}}(t) = \nu_{\text{eff}}(t_0){\left(1+\frac{A_z(t)}{r_s B_0 (\int  d \mathcal{V}  \, |\alpha_z(r)|^2)^{1/2} \gamma_c}\right)^{-2/3}}
\end{equation}

\begin{equation} \label{nu_eff_strong_shear}
\nu_{\text{eff}}(t) = \nu_{\text{eff}}(t_0) \left(\frac{r_s B_0 (\int  d \mathcal{V}  \, |\alpha_z(r)|^2)^{1/2} \gamma_c}{4A_z(t)}\right)^{1/3}
\end{equation}
in the weak and strong shear regimes, respectively. 

Starting with the weak-shear case, we take the time derivative of Eq.~(\ref{nu_eff_weak_shear}). Note that, in our model, the only time dependence comes from the zonal flow amplitude $A_z(t)$, which due to the beat-driven assumption, grows at exactly twice the rate of the pump wave. This results in $dA_z/dt \sim 2\left(\gamma_L - \gamma_{d,p}\right)A_z$, which allows us to simplify the derivative to the following form:

\begin{equation}
\frac{d \nu_{\text{eff}}(t)}{dt} = -\frac{\nu_{\text{eff}}^{5/2}(t)}{\nu_{\text{eff}}(t_0)^{3/2}}  \frac{4A_z(\gamma_L - \gamma_{d,p})}{3r_s B_0 (\int  d \mathcal{V}  \, |\alpha_z(r)|^2)^{1/2} \gamma_c}
\end{equation}
For the weak-shear case, we have that the shear induced by the ZM, defined in Eq.~(\ref{gamma_E}) as $\gamma_E = A_z(t)/r_s B_0(\int  d \mathcal{V}  \, |\alpha_z(r)|^2)^{1/2}$, is at most $\gamma_c$. It follows from Eq.~(\ref{nu_eff_weak_shear}) that:

\begin{equation}
\frac{\nu_{\text{eff}}(t)}{\nu_{\text{eff}}(t_0)}\sim \mathcal{O}(1) 
\end{equation}
Therefore, we have that 

\begin{equation}
\frac{1}{\nu_{\text{eff}}^2(t)}\frac{d \nu_{\text{eff}}(t)}{dt} \sim \frac{\gamma_L - \gamma_{d,p}}{\nu_{\text{eff}}(t)}.
\end{equation}

In the time-local regime, where condition (\ref{app:timelocal}) is valid, then the adiabaticity condition for the weak-shear case is automatically satisfied.

In the strong shear case we follow a similar procedure. Using again that $dA_z/dt \sim 2\left(\gamma_L - \gamma_{d,p}\right)A_z$ for the beat-driven ZM, together with Eq.~(\ref{nu_eff_strong_shear}) we obtain:

\begin{equation}
\frac{d \nu_{\text{eff}}(t)}{dt} = -\frac{2}{3} \left(\gamma_L - \gamma_{d,p}\right) \nu_{\text{eff}}(t),
\end{equation}
that, as for the weak-shear case, links the adiabaticity condition to the time-local regime:

\begin{equation}
\frac{1}{\nu_{\text{eff}}^2(t)}\frac{d \nu_{\text{eff}}(t)}{dt} \sim \frac{\gamma_L - \gamma_{d,p}}{\nu_{\text{eff}}(t)}
\end{equation}

Again, if the condition (\ref{app:timelocal}) is valid, then the adiabaticity condition is satisfied and the effect of the time-dependent scattering rate is time-local.

\section*{References}

%\bibliography{references}

\providecommand{\noopsort}[1]{}\providecommand{\singleletter}[1]{#1}%
\providecommand{\newblock}{}

\end{document}